\documentclass[twocolumn]{aastex61}
\usepackage{amsmath}
\usepackage{rotating}
\usepackage[strict]{changepage}
\usepackage{tabularx}

\hypersetup{linkcolor=red,citecolor=blue,filecolor=cyan,urlcolor=magenta}

\submitjournal{ApJ}

\shorttitle{coronal wave interaction with coronal holes}
\shortauthors{Piantschitsch et al.}

\begin{document}

\newcommand{\etal}{{\it et~al.}}
\newcommand{\ie}{{\it i.e.}}
\newcommand{\eg}{{\it e.g.}}

\newcommand{\goes}{{\it GOES}}
\newcommand{\sdo}{{\it SDO}}​

\title{Numerical Simulation of coronal waves interacting with coronal holes: \\ II. Dependence on Alfv\'{e}n speed inside the coronal hole}

\correspondingauthor{Isabell Piantschitsch}
\email{isabell.piantschitsch@uni-graz.at}

\author[0000-0002-0786-7307]{Isabell Piantschitsch}
\affiliation{IGAM/Institute of Physics, University of Graz, Universit\"atsplatz 5, A-8010 Graz, Austria}

\author{Bojan Vr\v{s}nak}
\affiliation{Hvar Observatory, Faculty of Geodesy, Ka\v{c}i\'{c}eva 26, HR-10000 Zagreb, Croatia}

\author{Arnold Hanslmeier}
\affiliation{IGAM/Institute of Physics, University of Graz, Universit\"atsplatz 5, A-8010 Graz, Austria}

\author{Birgit Lemmerer}
\affiliation{IGAM/Institute of Physics, University of Graz, Universit\"atsplatz 5, A-8010 Graz, Austria}

\author{Astrid Veronig}
\affiliation{IGAM/Institute of Physics, University of Graz, Universit\"atsplatz 5, A-8010 Graz, Austria}

\author{Aaron Hernandez-Perez}
\affiliation{IGAM/Institute of Physics, University of Graz, Universit\"atsplatz 5, A-8010 Graz, Austria}

\author{Ja\v{s}a \v{C}alogovi\'{c}}
\affiliation{Hvar Observatory, Faculty of Geodesy, Ka\v{c}i\'{c}eva 26, HR-10000 Zagreb, Croatia}

\begin{abstract}

We used our newly developed magnetohydrodynamic (MHD) code to perform 2.5D simulations of a fast-mode MHD wave interacting with coronal holes (CH) of varying Alfv\'{e}n speed which result from assuming different CH densities. We find that this interaction leads to effects like reflection, transmission, stationary fronts at the CH boundary and the formation of a density depletion that moves in the opposite direction to the incoming wave. We compare these effects with regard to the different CH densities and present a comprehensive analysis of morphology and kinematics of the associated secondary waves. We find that the density value inside the CH influences the phase speed as well as the amplitude values of density and magnetic field for all different secondary waves. Moreover, we observe a correlation between the CH density and the peak values of the stationary fronts at the CH boundary. The findings of reflection and transmission on the one hand and the formation of stationary fronts caused by the interaction of MHD waves with CHs on the other hand, strongly support the theory that large scale disturbances in the corona are fast-mode MHD waves. 
\end{abstract}


\section{Introduction} \label{sec:intro}

Large-scale propagating disturbances in the corona or coronal waves, as they are also called, were directly observed for the first time by the Extreme-ultraviolet Imaging Telescope (EIT; \citealt{Delaboudiniere1995}) onboard the Solar and Heliospheric Observatory (SOHO; \citealt{DomingoFleck1995}). They are driven by solar flares or alternatively by coronal mass ejections (CMEs) (for a comprehensive review see, \eg,\citealt{Vrsnak_Cliver2008}) and can be observed over the entire solar surface.

Inconsistencies regarding the analysis and comparison of observations and simulations led to the development of different theories on how to interpret coronal waves \citep{Long2017}. Coronal waves can either be described by wave theories, which consider the disturbances as fast-mode MHD waves  \citep{Vrsnak_Lulic2000,Lulic_etal2013,Warmuth2004,Veronig2010,Thompson1998,Wang2000,Wu2001,Ofman2002,Patsourakos2009,Patsourakos_etal.2009,Schmidt_Ofman2010}. Alternatively, coronal waves can be explained by so called pseudo-wave theories, which consider the observed disturbances as a result of the reconfiguration of the coronal magnetic field, caused by either continuous small-scale reconnection \citep{Attrill2007a,Attrill2007b,van_Driel-Gesztelyi_etal_2008}, Joule heating \citep{Delanee_Hochedez2007} or stretching of magnetic field lines \citep{Chen_etal2002}. Effects like reflection, refraction or transmission of coronal waves at a coronal hole (CH) boundary support the wave theory whereas the existence of stationary bright fronts was one of the primary reasons for the development of the competing pseudo-wave theory. Another, alternative, approach is hybrid models which try to combine both wave and pseudo-wave theories by interpreting the outer envelope of a CME as a pseudo-wave which is followed by a freely propagating fast-mode-MHD wave \citep{Chen_etal2002,Chen_etal2005,Zhukov_Auchere2004,Cohen_etal2009,Chen_Wu2011,Downs_etal2011,Cheng_etal2012,Liu_Nitta_etal2010}.  Recent observations also include both the wave and non-wave approach into the interpretation of
an individual EUV wave event \citep{Zong2017}.

However, among these competing theories the wave interpretation is regarded as the best supported approach \citep{Long2017,Warmuth2015}. Observational evidence for the wave character of these large scale propagating disturbances is given by various authors who report about waves being reflected and refracted at a CH \citep{Kienreich_etal2012,Veronig_etal2008,Long_etal2008,Gopalswamy_etal2009} or waves being transmitted through a CH \citep{Olmedo2012} or EIT wave fronts pushing plasma downwards \citep{Veronig2011,Harra2011}, which is also consistent with the interpretation that EIT waves are fast-mode MHD waves. Recent observations also show that fast EUV waves are able to form bright stationary fronts at the boundary of a magnetic separatrix layer \citep{Chandra2016}. Furthermore, studies on simulating coronal waves indicate, that stationary wave fronts at a CH boundary can be produced by the interaction of a fast-mode MHD wave with obstacles like a CH \citep{Piantschitsch2017} or a magnetic quasi-separatrix layer \citep{Chen2016} and therefore confirm the above mentioned observations.

In \citet{Piantschitsch2017} we used a newly developed MHD code to perform 2.5D simulations which showed that the interaction of an MHD wave with a low density region like a CH leads to effects like reflection and  transmission of the incoming wave. Moreover, we observed stationary features at the CH boundary and the formation of a density depletion which is moving in opposite direction of the incoming wave propagation. We found reflections inside the CH which subsequently led to additional transmissive and reflective features outside the CH. We showed that the incoming wave pushes the CH boundary in the direction of wave propagation. Additionally, we compared phase speeds and positions of the incoming wave and the resulting waves after the interaction with a CH and found good agreement with observational cases where waves were being reflected and refracted at a CH \citep{Kienreich_etal2012} or transmitted through a CH \citep{Olmedo2012}.

In \citet{Piantschitsch2017} we assumed a certain initial density amplitude for the incoming wave and a fixed CH density for our simulations. In this paper we focus on the comparison of different CH densities and on how these various densities change the kinematics of the secondary waves (i.e. reflected, transmitted and traversing waves) and the stationary features at the CH boundary. These different CH densities lead subsequently to different Alfv\'{e}n speeds inside the CH. We will show that there is an influence of the CH density and the Alfv\'{e}n speed, respectively, on the amplitude values of the secondary waves and the peak values of the stationary features.

In Section 2 we describe the initial conditions and present the numerical method we use for our simulations. A comprehensive description of the morphology of the reflected, traversing and transmitted waves will be presented in Section 3. In Section 4 we analyze the kinematic measurements of secondary waves and stationary features and compare the cases of varying CH densities. In Section 5 and Section 6 we discuss the conclusions that can be drawn from our simulation results.

\section{Numerical setup}
\subsection{Algorithm and Equations}

We use our newly developed code to perform 2.5D simulations of MHD wave propagation and its interaction with low density regions of varying density. In this code we numerically solve the standard homogeneous MHD equations (for detailed description of the equations see \citet{Piantschitsch2017}) by applying the so called Total Variation Diminishing Lax-Friedrichs (TVDLF) scheme, first described by \citet{Toth_Odstrcil1996}. This scheme is a fully explicit method and achieves second order accuracy in space and time. The simulations are performed by using a $500\times300$ resolution and a dimensionless length of the computational box equal to $1.0$ both in the $x$- and $y$-direction. Transmissive boundary conditions are used for the simulation boundaries.

\subsection{Initial Conditions}

We assume an idealized case with zero pressure all over the computational box and a homogeneous magnetic field in the vertical direction. The initial setup describes five different cases for the density distribution inside the CH, starting from a density value of $\rho_{CH}=0.1$ and going up to $\rho_{CH}=0.5$. The detailed initial conditions for all parameters are as follows:

\begin{equation}
     \rho(x) = 
    \begin{cases}
        \Delta\rho\cdot cos^2(\pi\frac{x-x_0}{\Delta x})+\rho_0 & 0.05\leq x\leq0.15 \\
        0.1 \lor 0.2 \lor 0.3 \lor 0.4 \lor 0.5 & \:\:0.4\leq x\leq0.6 \\
        \qquad \qquad1.0 & \:\:\qquad\text{else}
    \end{cases}
\end{equation}

\begin{equation}
    v_x(x) = 
    \begin{cases}
        2\cdot \sqrt{\frac{\rho(x)}{\rho_0}} -2.0& 0.05\leq x\leq0.15 \\
        \:\:0 & \:\:\qquad\text{else}
    \end{cases}
\end{equation}

\begin{equation}
    B_z(x) = 
    \begin{cases}
        \:\:\rho(x) & 0.05\leq x\leq0.15 \\
        \:\: 1.0 & \:\:\qquad\text{else}
    \end{cases}
\end{equation}

\begin{equation}
B_{x}=B_{y}=0,\qquad0\leq x\leq1
\end{equation}

\begin{equation}
v_{y}=v_{z}=0,\qquad0\leq x\leq1
\end{equation}

where $\rho_{0}=1.0$, $\triangle\rho=0.5$,
$x_{0}=0.1$, $\triangle x=0.1$.

\begin{figure}[ht!]
\centering\includegraphics[width=0.48\textwidth]{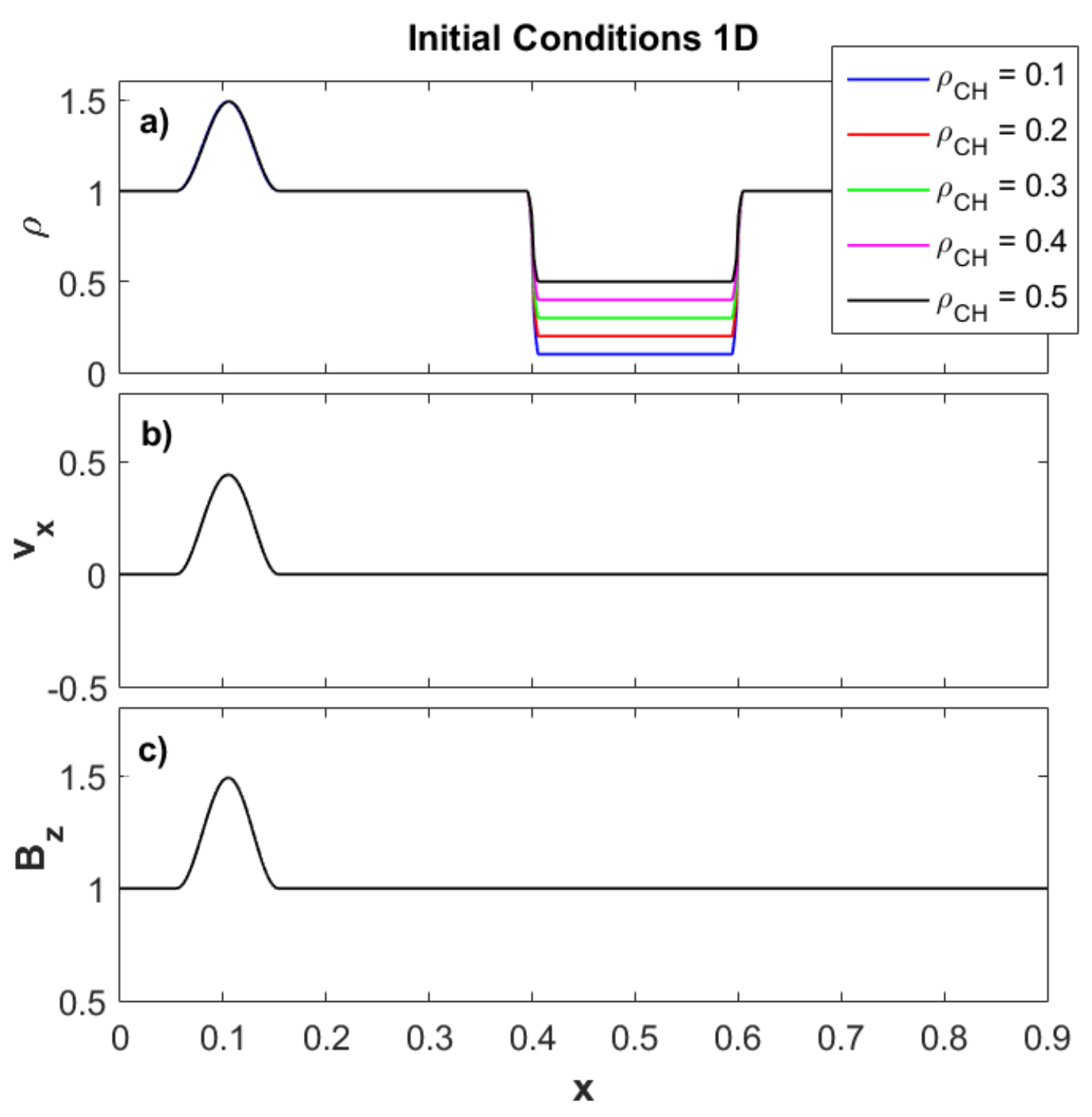}
\caption{Initial conditions for density $\rho$, plasma flow velocity $v_x$ and magnetic field in $z$-direction $B_z$ for five different densities inside the CH, starting from $\rho_{CH}=0.1$ (blue solid line), increased by steps of $0.1$ and ending with $\rho_{CH}=0.5$ (black solid line in the range $0.4\leq x\leq0.6$).}
\label{InitCond_1D}
\end{figure}

\begin{figure*}[ht]
\centering \includegraphics[width=\textwidth]{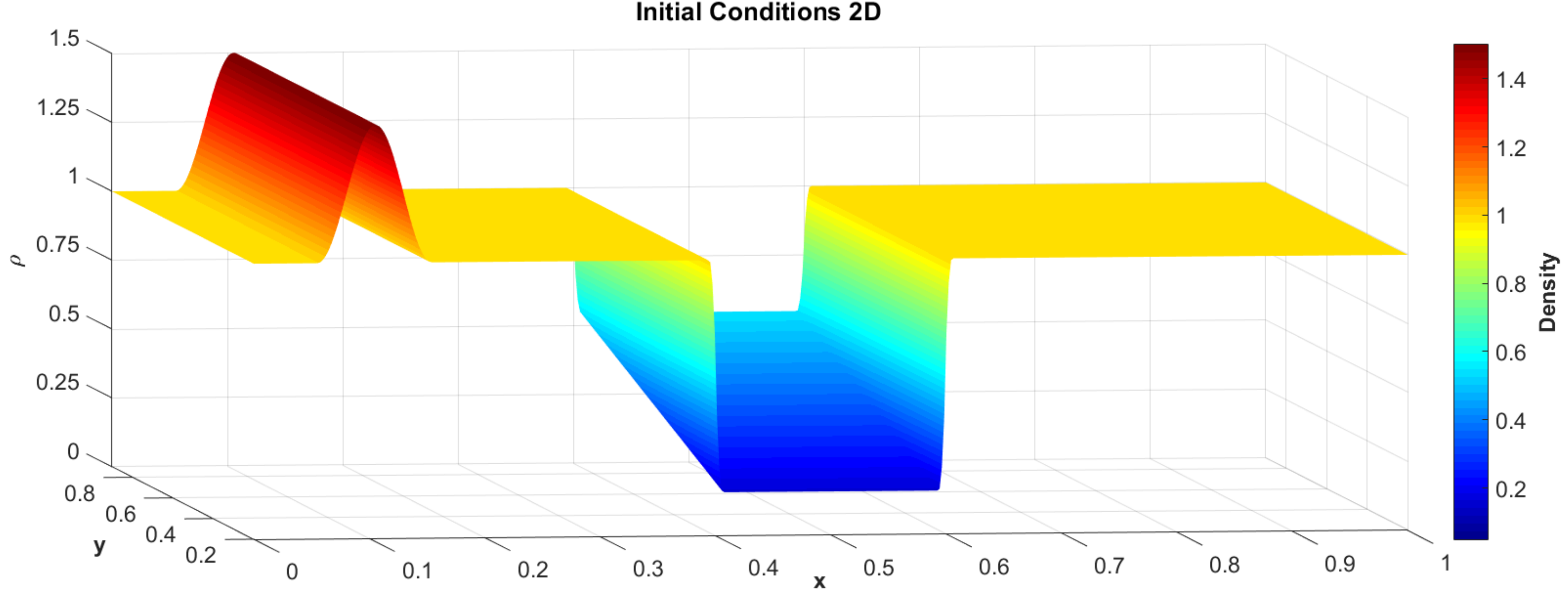}
\caption{Initial two-dimensional density distribution, showing a fixed initial wave amplitude of $\rho=1.5$ and a linearly increasing CH density from $\rho_{CH}=0.1$ up to $\rho_{CH}=0.5$ in the range $0.4\leq x\leq0.6$. The background density is equal to one. }
\label{Initcond_2D}
\end{figure*}

\begin{figure*}[!htb]
\centering \includegraphics[width=\textwidth,height=23cm]{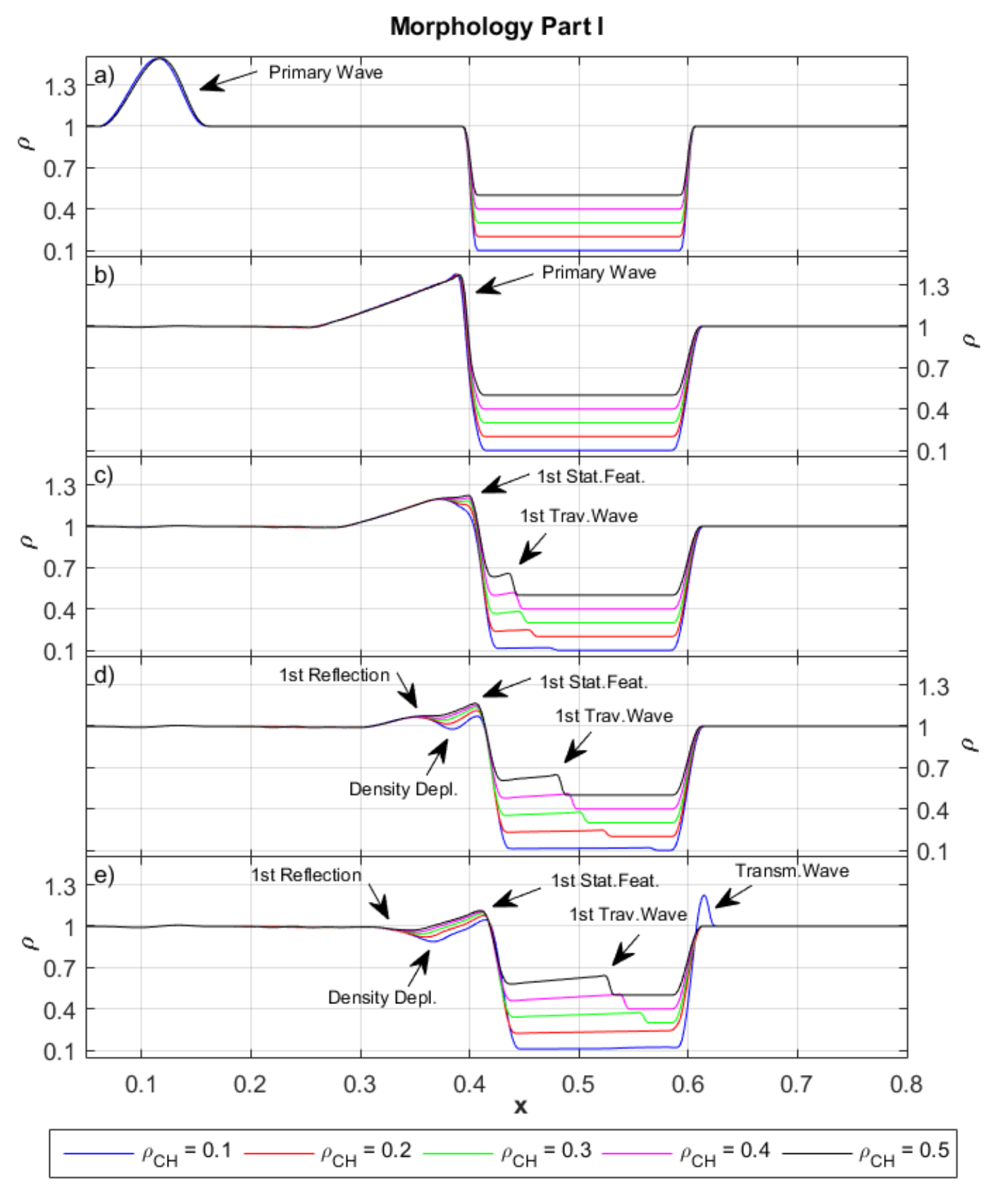}
\caption{Overlay of the temporal evolution of the density distribution for all five different CH densities. Starting at the beginning of the simulation run at $t=0$ (panel a) and ending when the first transmitted wave occurs at $x\approx0.6$ (blue line in panel e). The arrows denote the position of the primary wave, the first reflection, the first stationary feature and the first traversering wave (for the case $\rho_{CH}=0.5$, black) as well as the density depletion and the transmitted wave (for the case $\rho_{CH}=0.1$, blue). (An animation of this figure is available in the online journal.)}
\label{morphology_1D_part1}
\end{figure*}

\begin{figure*}[!htb]
\centering \includegraphics[width=\textwidth,height=23cm]{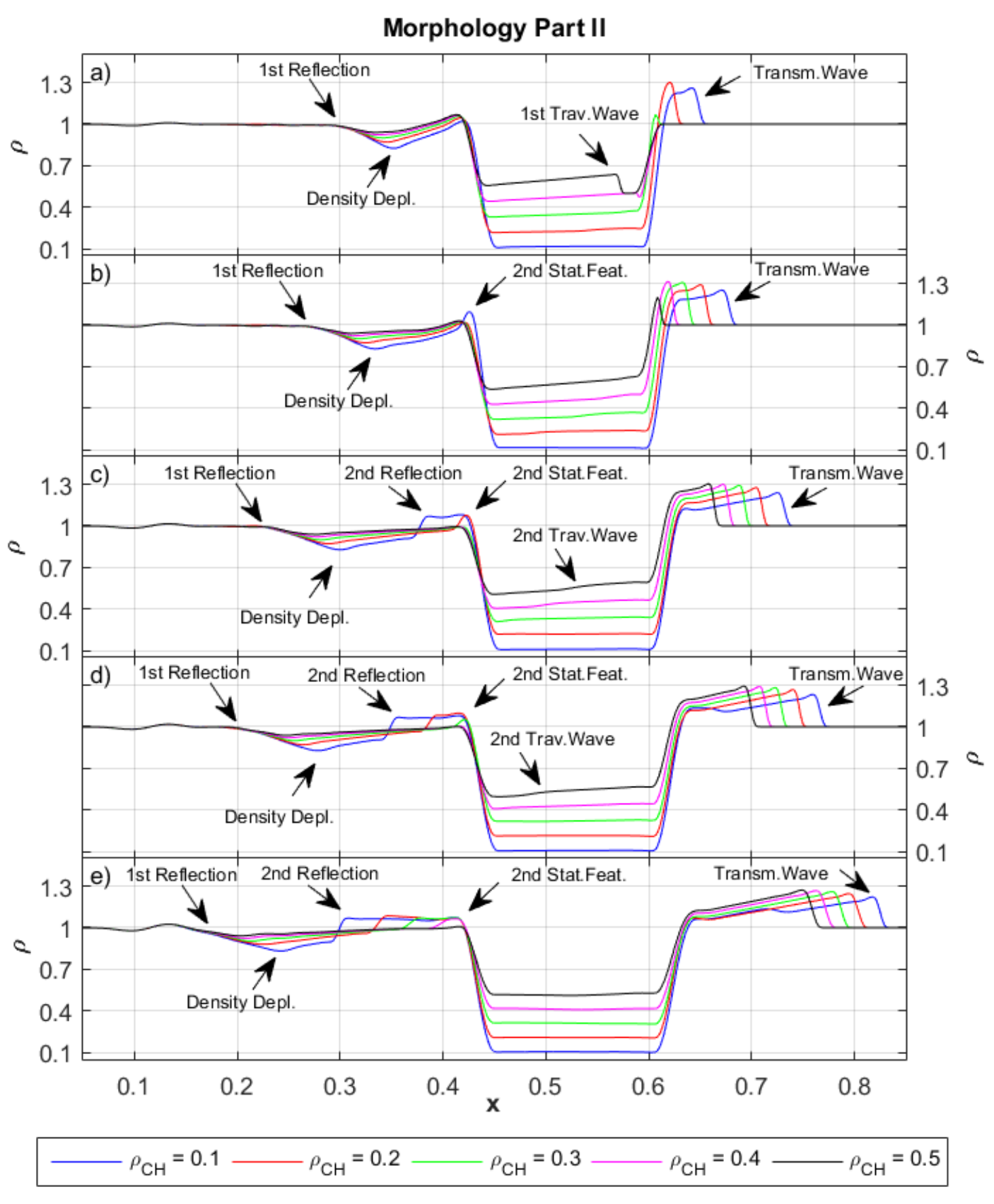}
\caption{Continuation of Figure \ref{morphology_1D_part1}. Overlay of the temporal evolution of the density distribution for all five different CH densities. Starting when the second transmitted wave occurs at $x\approx0.6$ in panel a) (red peak) and ending at the end of the simulation run at $t=0.5$ (panel e)). The arrows denote the position of the first reflection, the first and the second traversing wave (for the case $\rho_{CH}=0.5$, black) as well as the position for the density depletion, the second reflection and the transmitted wave (for the case $\rho_{CH}=0.1$, blue). The first occurance of the second stationary feature is shown in the cases of four different CH densities (blue, red, green, magenta) in the panels b-e. (An animation of this figure is available in the online journal.)}
\label{morphology_1D_part2}
\end{figure*}

Figure \ref{InitCond_1D} shows a vertical cut through the 2D initial conditions for density, $\rho$, $z$-component of the magnetic field, $B_{z}$, and plasma flow velocity in $x$-direction, $v_{x}$. In Figure \ref{InitCond_1D}a we see an overlay of five different vertical cuts of the 2D density distribution at $y=0$ ($\rho_{CH}=0.1$), $y=0.25$ ($\rho_{CH}=0.2$), $y=0.5$ ($\rho_{CH}=0.3$), $y=0.75$ ($\rho_{CH}=0.4$) and $y=1$ ($\rho_{CH}=0.5$). In the range $0.05\leq x \leq 0.15$ we created a wave with the initial amplitude of $\rho=1.5$ (for detailed description see equation (1)). We can see that the initial density amplitude of the incoming wave has the same value in all five cases whereas the initial CH densities within the range $0.4\leq x \leq 0.6$ vary from one vertical cut to another. The background density is equal to 1.0 everywhere. Figure \ref{InitCond_1D}b and Figure \ref{InitCond_1D}c show the initial conditions for plasma flow velocity in $x$-direction, $v_{x}$, and $z$-component of the magnetic field, $B_{z}$. We can see that $B_{z}$ and $v_{x}$ are defined as functions of $\rho$ in the range $0.05\leq x \leq 0.15$. The initial amplitudes for $v_{x}$ and $B_{z}$ are the same for all cases of varying density distribution. The background magnetic field in the $z$-direction is equal to $1.0$ over the whole computational box whereas the magnetic field components in $x$- and $y$-direction are equal to zero everywhere (see Equations (3) and (4)). The background plasma flow velocity in the $x$-direction is equal to zero and the plasma flow velocities for the $y$- and $z$-directions are equal to zero over the whole computational grid.

In Figure \ref{Initcond_2D} we see the 2D initial conditions for the density distribution, showing a linearly increasing density from $\rho_{CH}=0.1$ up to $\rho_{CH}=0.5$ in the range $0.4\leq x \leq 0.6$. The initial density amplitude of the incoming wave has the same value along the whole $y$-axis. This initial 2D setup enables us to perform simulations of the wave propagation for different CH densities simultaneouosly. Hence, the differences of phase speed and amplitude values due to varying densities inside the CH can be compared immediately.

\section{Morphology}

In Figure \ref{morphology_1D_part1} and Figure \ref{morphology_1D_part2} we have plotted the temporal evolution of the density distribution for five different CH densities, starting at the beginning of the simulation run at $t=0$ and ending at $t=0.5$. We can see overlayed vertical cuts through the $xz$-plane of our simulations at $y=0$, $y=0.25$, $y=0.5$, $y=0.75$ and $y=1$ at ten different time steps. We observe the temporal evolution of the incoming wave (hereafter named primary wave) and its interaction with the CHs of different density values. Moreover we can see the different behaviour of the reflected, transmitted and traversing waves (hereafter named secondary waves) due to varying density values inside the CH. We observe different kinds of stationary effects at the left CH boundary and also density depletions of varying depths, moving in the negative $x$-direction. In addition to that we find that the primary wave is able to push the left CH boundary in the direction of the primary wave's propagation.

\subsection{Primary Wave}

In Figure \ref{morphology_1D_part1}a and Figure \ref{morphology_1D_part1}b we can see how the primary wave is moving in the positive $x$-direction towards the left CH boundary. At the same time when the density amplitude starts decreasing we observe a broadening of the width of the wave that is accompanied by a steepening of the wave and a subsequent shock formation.

\subsection{Secondary Waves}

After the primary wave has reached the left CH boundary (see Figure \ref{morphology_1D_part1}c) we find that the density amplitude quickly decreases when the wave starts traversing through the CH. The smallest density amplitude inside the CH can be seen for the case of an initial CH density of $\rho_{CH}=0.1$ (blue), whereas the largest wave amplitude is observed in the case of $\rho_{CH}=0.5$ (black). Furthermore, we observe immediate responses of the primary wave's impact on the left CH boundary (see Figure \ref{morphology_1D_part1}c and \ref{morphology_1D_part1}d). First, one can see a stationary feature which appears as a stationary peak at $x\approx0.4$ in Figures \ref{morphology_1D_part1}c-\ref{morphology_1D_part1}e and Figure \ref{morphology_1D_part2}a. The morphology of this stationary feature will be discussed in Section 3.3. Second, we observe a first reflective feature (seen at $x\approx0.37$ in Figures \ref{morphology_1D_part1}c and \ref{morphology_1D_part1}d), which is not able to move onward in the negative $x$-direction until the incoming wave has not completed the entry phase into the CH. Figure \ref{morphology_1D_part1}e and Figures \ref{morphology_1D_part2}a - \ref{morphology_1D_part2}d show how this first reflection is finally moving toward the negative $x$-direction, where it is then difficult to distinguish from the background density. The first reflection is the same in all five cases of varying initial density inside the CH and it is located at the left side of the density depletions. The smaller the initial density value inside the CH, the smaller the minimum value of the density depletion, \ie\ the stronger the density depletion.

\begin{figure*}[ht]
\centering \includegraphics[width=\textwidth]{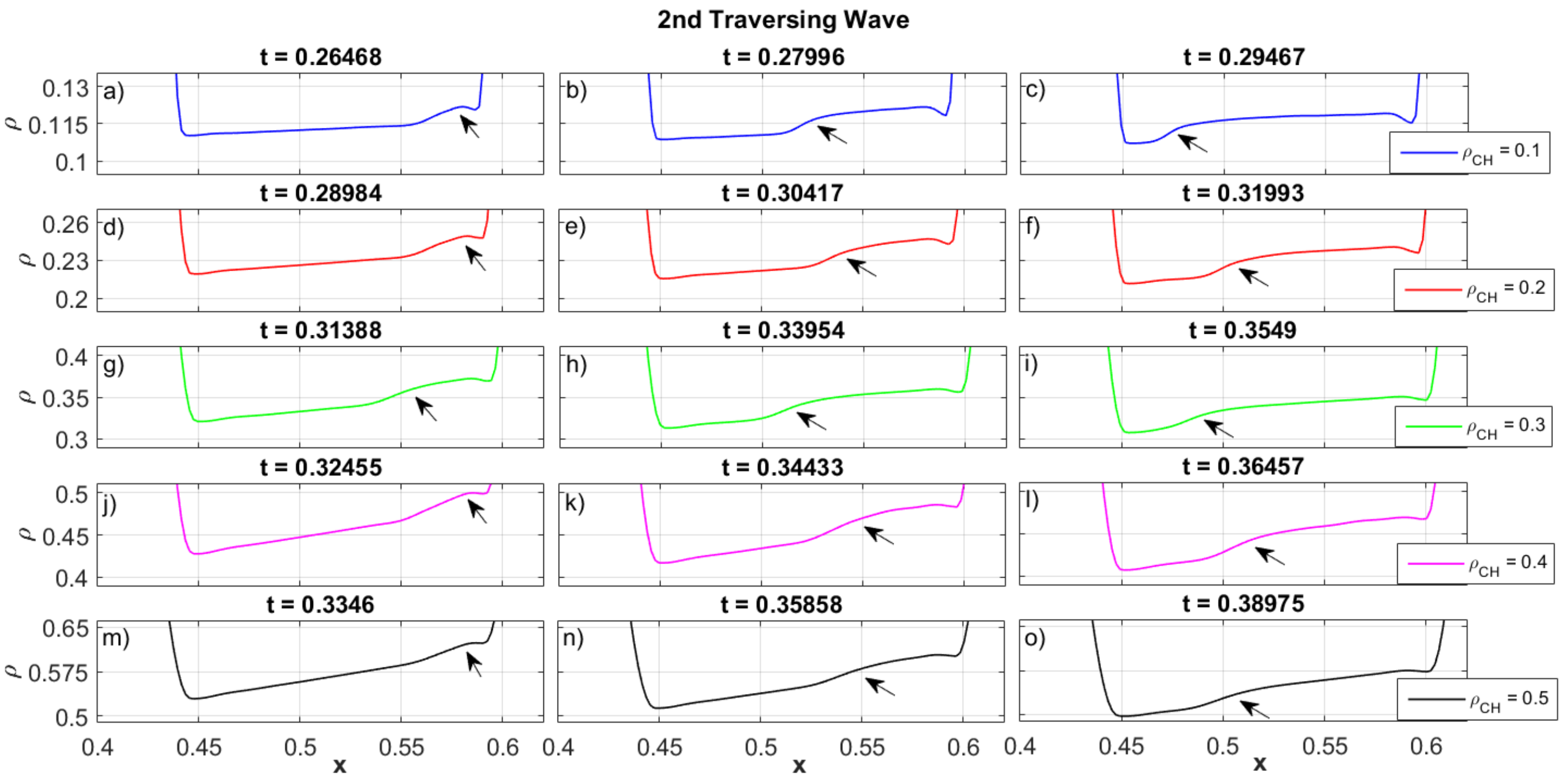}  
\caption{Temporal evolution of the density distribution of the second traversing wave moving in negative $x$-direction inside the CH. Starting shortly after the first traversing wave with the lowest amplitude (blue) got reflected at the right CH boundary inside the CH at $t=0.26468$ and ending before the second traversing wave with the largest amplitude (black) reaches the left CH boundary inside the CH ($t=0.38975$). The arrows denote the wave crest of the second traversing wave for all cases of different CH density.}
\label{morphology_IN_CH_no1}
\end{figure*}

\begin{figure*}[ht]
\centering \includegraphics[width=\textwidth]{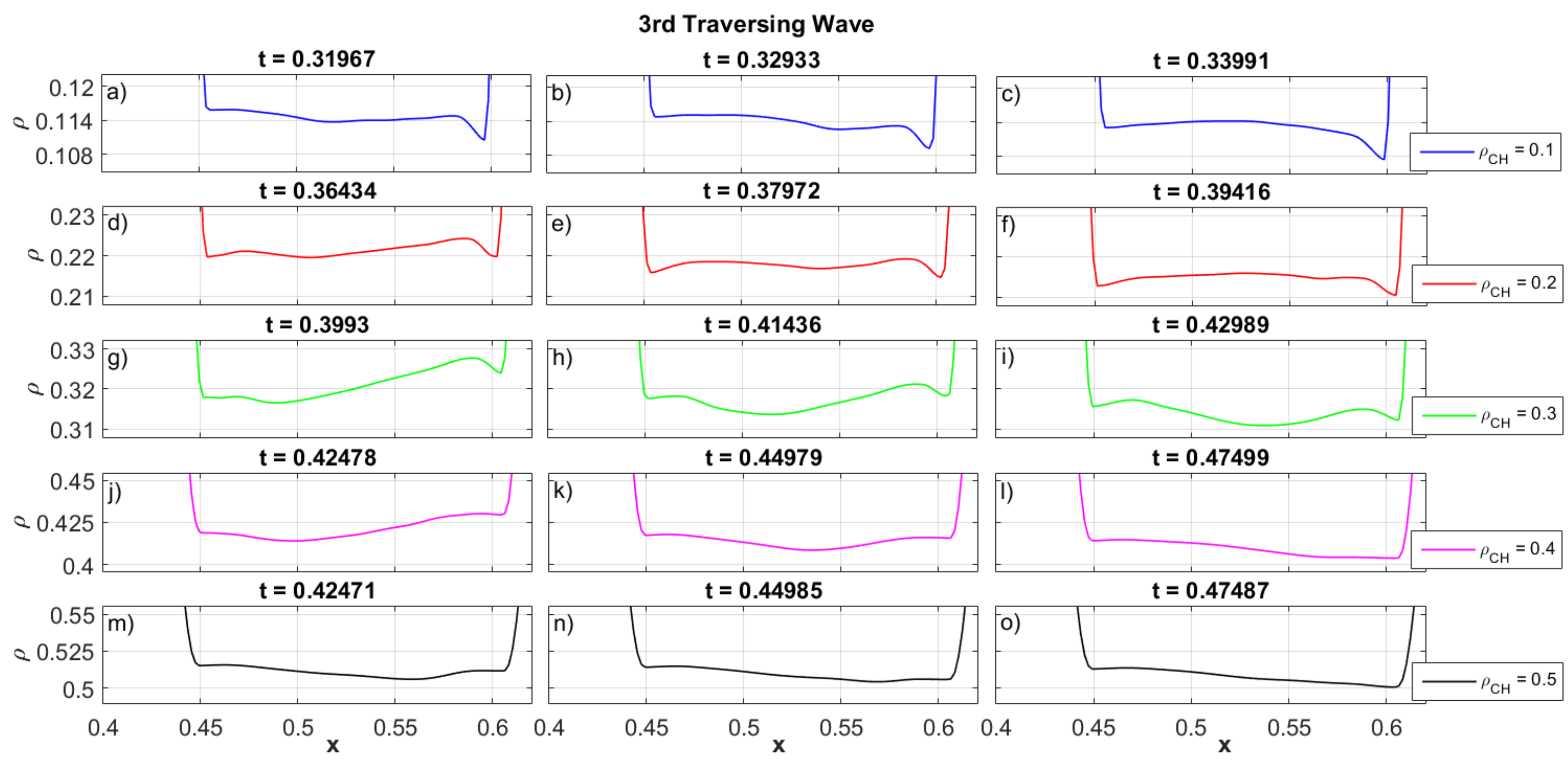}  
\caption{Temporal evolution of the density distribution of the third traversing wave moving in positive $x$-direction inside the CH. Starting shortly after the second traversing wave with the lowest amplitude (blue) got reflected at the left CH boundary inside the CH at $t=0.31967$ and ending before the third traversing wave with the largest amplitude (black) reaches the right CH boundary inside the CH ($t=0.47487$).}
\label{morphology_IN_CH_no2}
\end{figure*}

In Figures \ref{morphology_1D_part1}c - \ref{morphology_1D_part1}e one can see how the waves are traversing through the CH with much lower density amplitude than that of the primary wave. We observe that the smaller the initial density value inside the CH, the smaller the density amplitude of the traversing wave and the faster the wave propagates through the CH. For a better comparison of the different traversing waves we zoom in the region $0.4\leq x\leq0.6$ in Figure \ref{morphology_IN_CH_no0}, Figure \ref{morphology_IN_CH_no1} and Figure \ref{morphology_IN_CH_no2}. We choose the time interval from $t=0.22481$ to $t=0.47487$ which is the time interval where the traversing waves are moving back and forth inside the CH. Figure \ref{morphology_IN_CH_no0} shows that the waves are moving with approximately constant density amplitudes of $\rho=0.11$ (for $\rho_{CH}=0.1$), $\rho=0.25$ (for $\rho_{CH}=0.2$), $\rho=0.39$ (for $\rho_{CH}=0.3$), $\rho=0.52$ (for $\rho_{CH}=0.4$) and $\rho=0.65$ (for $\rho_{CH}=0.5$) towards the right CH boundary inside the CH. We saw in Figure \ref{morphology_1D_part2} that at the time when the traversing waves reach the right CH boundary, one part of each wave is leaving the CH and is propagating onwards as a transmitted wave. In Figure \ref{morphology_IN_CH_no1} we find that another part of the traversing waves gets reflected at the right CH boundary inside the CH. When these second traversing waves, which are propagating in the negative $x$-direction now, reach the left CH boundary, again one part leaves the CH hole and this causes another stationary feature at the CH boundary outside of the CH (seen as sharp peaks in Figure \ref{morphology_1D_part2}b - \ref{morphology_1D_part2}d), this will be discussed in Section 3.3. Every one of these stationary features is followed by a wave that is moving in the negative $x$-direction (second reflection) while the stationary features can still be observed. The density ampitudes of this second reflection do not have a clear correlation with the initial density values inside the CH. A detailed analysis of the parameters of the second reflection will be performed in Section 4.2.

Due to the varying phase speeds inside the CH the traversing waves leave the CH at different times. Hence, the smaller the density value inside the CH, the earlier we can observe the transmitted wave propagating outside of the CH. After leaving the CH, all different transmitted waves keep moving onwards in the positive $x$-direction until the end of the simulation run at $t=0.5$ (see Figures \ref{morphology_1D_part2}a - \ref{morphology_1D_part2}e). One can see that the smaller the initial density inside the CH, the smaller the density amplitude of the transmitted wave (smallest density amplitude for the transmitted wave in case of $\rho_{CH}=0.1$, marked in blue; largest density amplitude for the transmitted wave in case of $\rho_{CH}=0.5$, marked in black).

Besides causing a second reflection, the second traversing waves get reflected inside the CH again and move a third time through the CH, now again in the positive $x$-direction (see Figure \ref{morphology_IN_CH_no2}). When this third traversing wave reaches the right CH boundary, it causes a kind of subwave inside the first transmitted wave, seen as a peak inside the already existing transmitted wave in Figure \ref{morphology_IN_CH_no2}. Since we can not see these additional peaks inside the transmitted waves very clearly in Figure \ref{morphology_IN_CH_no2}, we zoom in the area of $0.6<x<0.9$. Figure \ref{zoom_transmitted_wave} shows these peaks at $x\approx0.76$ (for $\rho_{CH}=0.1$) and at $x\approx0.685$ (for $\rho_{CH}=0.2$). This kind of second transmission moves together with the first transmission in the positive $x$-direction until the end of the simulation run at $t=0.5$ but it can only be seen in the cases of $\rho_{CH}=0.1$ (blue) and $\rho_{CH}=0.2$ (red).

\subsection{Stationary Features}

We observe a first stationary feature at the left CH boundary, this appears as a stationary peak at $x\approx0.4$ in Figures \ref{morphology_1D_part1}c-\ref{morphology_1D_part1}e and Figure \ref{morphology_1D_part2}a. This peak occurs in all five cases of different initial CH density but it can be seen most clearly for the case of $\rho_{CH}=0.5$ (black line). During the lifetime of these stationary features the rear part of the primary wave continues entering the CH and the traversing waves keep moving onwards inside the CH. Due to the plot resolution on the one hand and the time delay of this feature for the different cases of $\rho_{CH}$ on the other hand, the single peaks are hard to distinguish and detect in Figures \ref{morphology_1D_part1}c and \ref{morphology_1D_part1}d. Hence, we zoom in the area $0.3\leq x\leq 0.5$ for the time period in which this first stationary feature occurs, in order to be able to study the peak values and the lifetime of this feature for all cases of different initial density inside the CH. Figure \ref{zoom_first_stat}    shows that a stationary peak can be observed first in the case of $\rho_{CH}=0.1$ (blue line) at $x\approx0.4$, followed by the peaks in the cases $\rho_{CH}=0.2$, $\rho_{CH}=0.3$, $\rho_{CH}=0.4$ and $\rho_{CH}=0.5$. The density of all peak values decreases in time, starting from $t=0.22481$ and ending at $t=0.29467$. One can see that the smaller the initial density inside the CH, the larger the peak value of this first stationary feature. Moreover, we can observe that the peak values move slighty in the positive $x$-direction for all different values of $\rho_{CH}$.

In Figures \ref{morphology_1D_part2}b, \ref{morphology_1D_part2}c and \ref{morphology_1D_part2}d we find a second stationary feature at the left CH boundary at about $x\approx0.43$. It occurs first for the case of $\rho_{CH}=0.1$ (blue), followed by the cases $\rho_{CH}=0.2$ (red), $\rho_{CH}=0.3$ (green), $\rho_{CH}=0.4$ (magenta) and $\rho_{CH}=0.5$ (black). In order to study the life time and the density peak values in detail we zoom in the region $0.3\leq x\leq 0.5$ between $t=0.31488$ and $t=0.44975$. In contrary to the first stationary feature, Figure \ref{zoom_second_stat} shows that the smaller the initial density inside the CH, the larger the peak value of the second stationary feature. When we compare the time when the peaks show up with the time evolution of the traversing wave inside the CH, we see that the second stationary features appear at the time when the second traversing waves have reached the left CH boundary inside the CH again. We also find that the smaller the initial $\rho_{CH}$ inside the CH is, the longer the lifetime of the second stationary peak. The peaks of this feature remain observable while the second reflection is moving onward in the negative $x$-direction (see Figures \ref{morphology_1D_part2}c,  \ref{morphology_1D_part2}d and  \ref{morphology_1D_part2}e).

\subsection{Density Depletion}

\begin{figure*}[ht]
\centering \includegraphics[width=\textwidth]{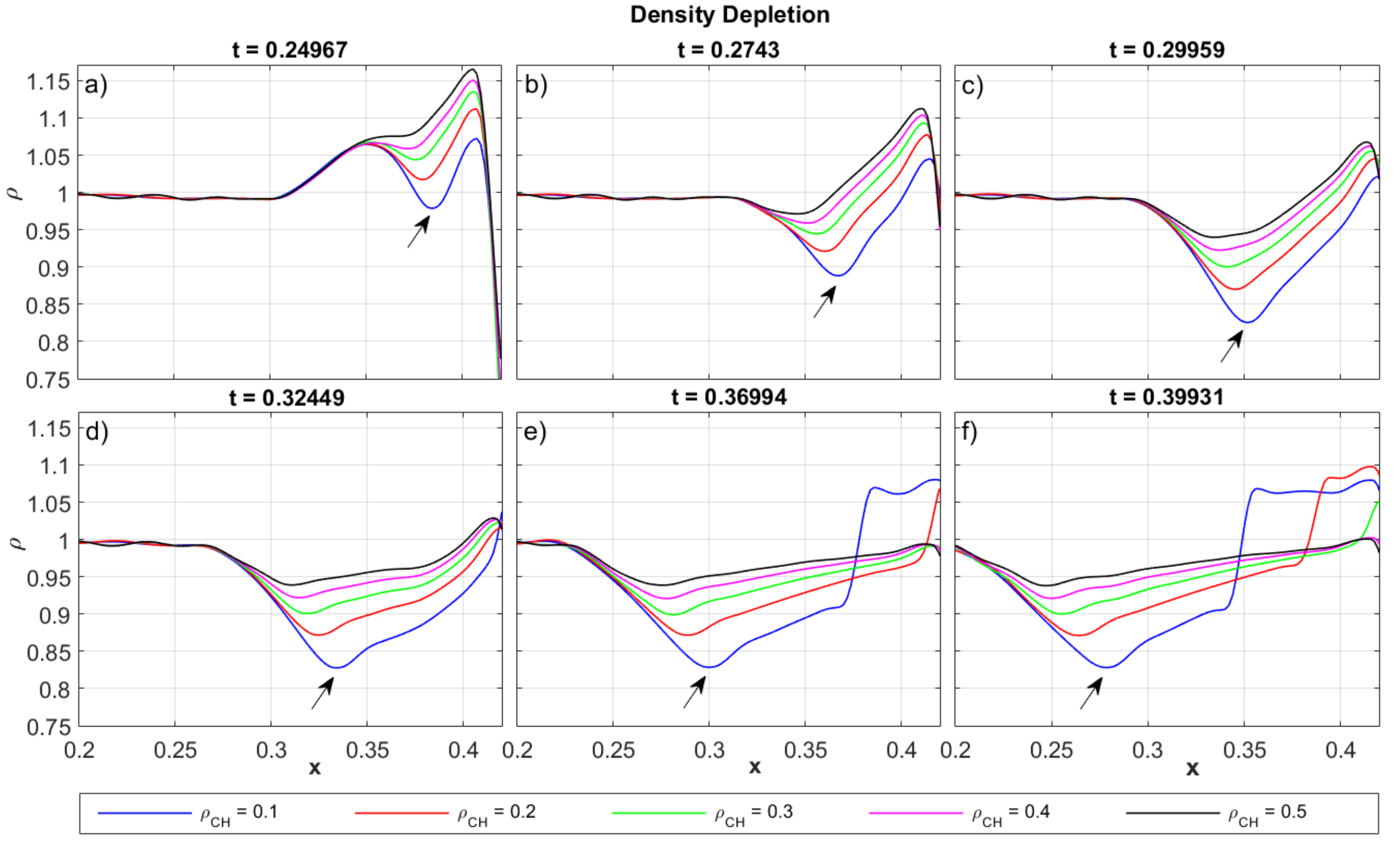}  
\caption{Zoom into the area of the density depletion ahead of the first reflection in the range $0.2\leq x\leq0.42$ for the cases $\rho_{CH}=0.1$ (blue), $\rho_{CH}=0.2$ (red), $\rho_{CH}=0.3$ (green), $\rho_{CH}=0.4$ (magenta) and $\rho_{CH}=0.5$ (black). The arrows denote the position of the minimum value of the density depletion for the case $\rho_{CH}=0.1$ (blue).}
\label{zoom_density_depletion}
\end{figure*}

In Figure \ref{morphology_1D_part1}d we observe the beginning of the evolution of a density depletion at $t\approx0.39$, most clearly seen for the case $\rho_{CH}=0.1$ (blue) and located at the left side of the first stationary feature. This density depletion appears for all values of $\rho_{CH}$ but has different minimum values. It propagates in the negative $x$-direction, ahead of the second reflection. One can see that the smaller the value for $\rho_{CH}$, the smaller is the minimum value of the density depletion. One more time we zoom in the area of interest ($0.2\leq x\leq0.4$) to analyze and compare the different depletions in detail. Figure \ref{zoom_density_depletion} shows the time evolution of the density depletions from $t=0.24967$ to $t=0.39931$.

\subsection{2D Morphology}

\begin{figure*}[ht!]
\centering \includegraphics[width=\textwidth]{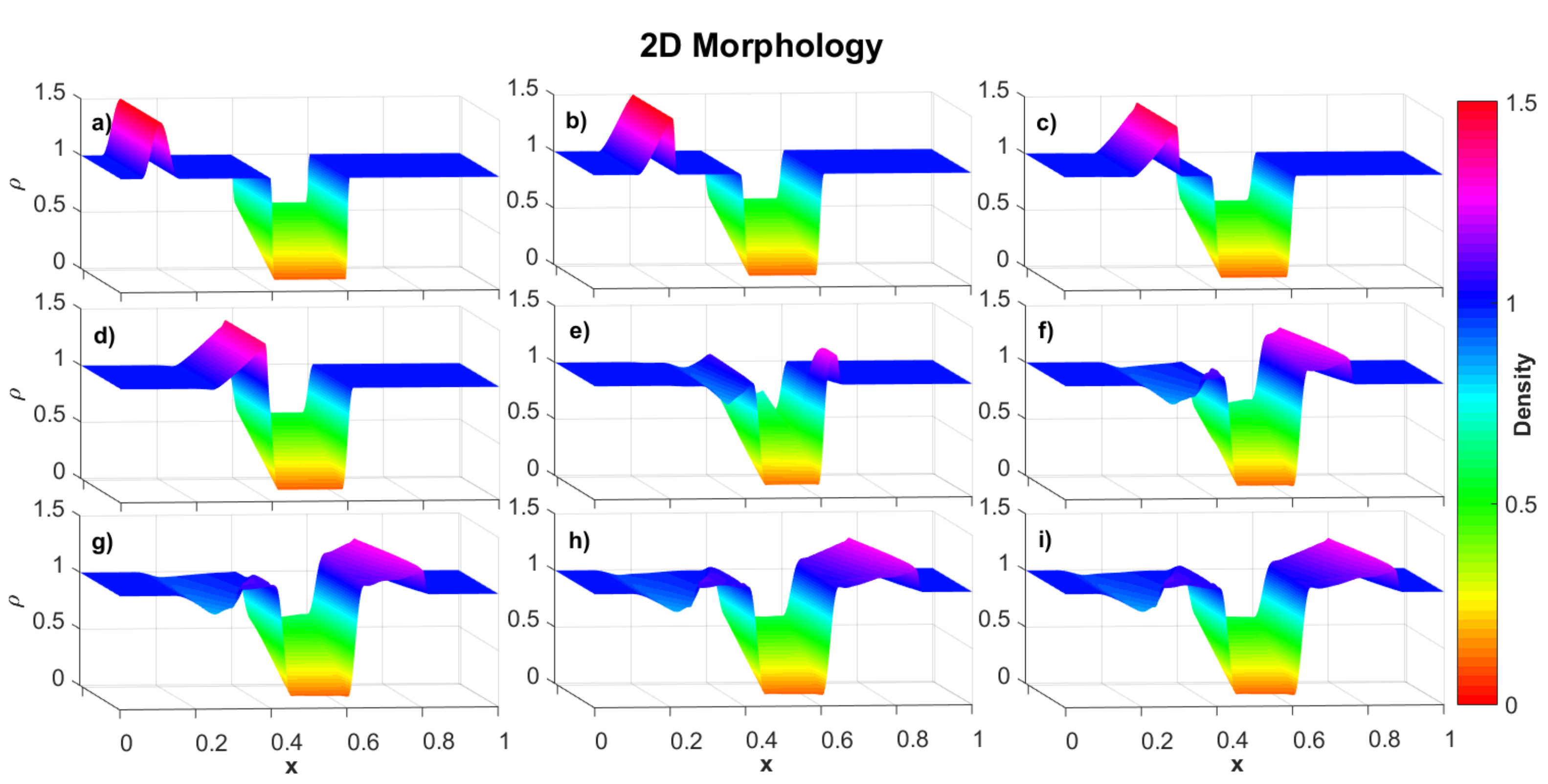}
\caption{Temporal evolution of density distribution for all different CH densities at the same time. Starting at the beginning of the simulation run at $t=0$ and ending at $t=0.5$.}
\label{morphology_rho_2D}
\end{figure*}

In Figure \ref{morphology_rho_2D} we see the 2D temporal evolution of the density distribution for all cases of varying $\rho_{CH}$, starting at $y=0$ with $\rho_{CH}=0.1$ and increasing linearly to $\rho_{CH}=0.5$ at $y=1$. Figure \ref{morphology_rho_2D}a shows the initial setup of the simulation run at $t=0$. In Figure \ref{morphology_rho_2D}b one can see the primary wave shortly before the entry phase into the CH. What we observe in Figures \ref{morphology_rho_2D}c and \ref{morphology_rho_2D}d is how the wave enters the CH and starts traversing through the CH. We find that the smaller the value of $\rho_{CH}$, the faster the wave traverses through the CH. Figure \ref{morphology_rho_2D}e shows that those waves which crossed a CH of low density already left the CH, while those which having entered a CH of a higher density value are still traversing through the CH. 
In Figure \ref{morphology_rho_2D}f we see that the second stationary feature starts appearing at the left CH boundary, caused by the first traversing waves reaching the left CH boundary. Moreover, in Figures \ref{morphology_rho_2D}g, \ref{morphology_rho_2D}h and \ref{morphology_rho_2D}i we can observe the evolution of the second stationary features for all cases of different $\rho_{CH}$ as well as the propagation of the density depletion and the second reflection in the negative $x$-direction. How the transmissive waves for all different values for $\rho_{CH}$ are moving forward in the positive $x$-direction can be seen in Figures \ref{morphology_rho_2D}e - \ref{morphology_rho_2D}i.

\begin{figure*}[ht]
\centering \includegraphics[width=\textwidth]{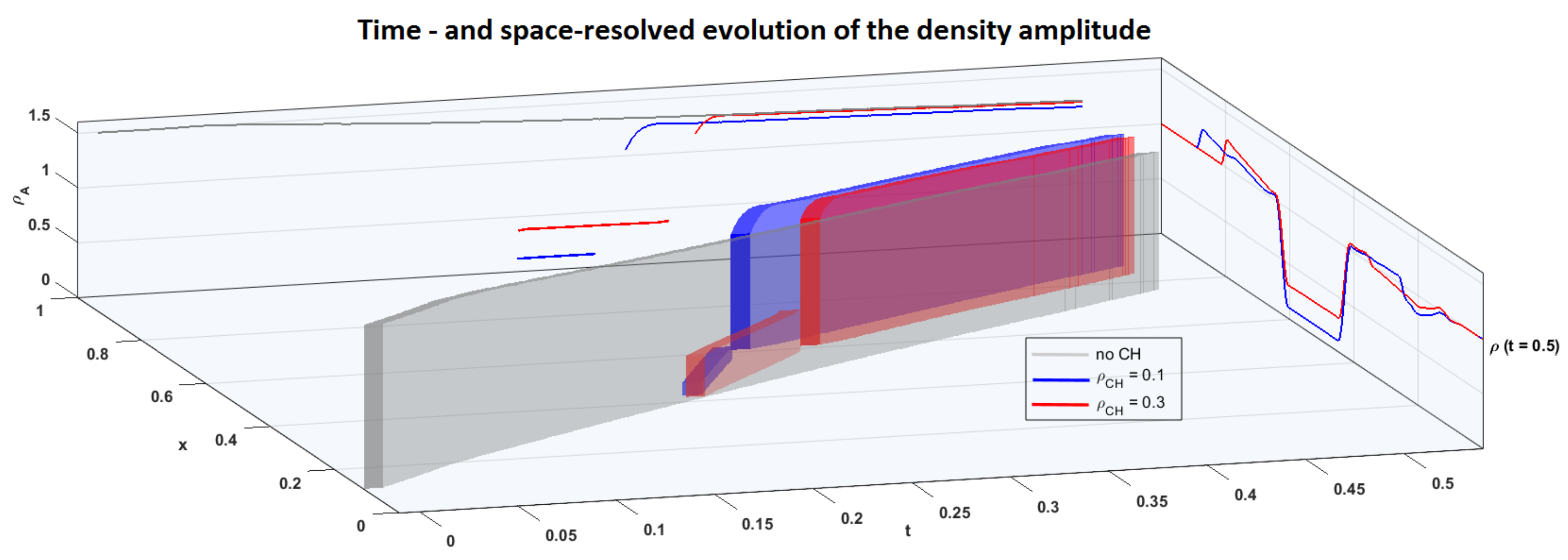}
\caption{The 3D plot in the center of the figure shows the temporal evolution of the amplitude position and the density amplitude $\rho_A$ for the cases $\rho_{CH}=0.1$ (blue) and $\rho_{CH}=0.3$ (red). The plot in the back compares the density amplitudes of the traversing and the transmitted wave for the cases $\rho_{CH}=0.1$ (blue) and $\rho_{CH}=0.3$ (red) with regard to the density amplitude of a wave with no interaction with a CH (gray). The plot on the right compares the density distribution at the end of the simulation run ($t=0.5$) for the cases $\rho_{CH}=0.1$ (blue) and $\rho_{CH}=0.3$ (red). }
\label{evolution_amplitude}
\end{figure*}

Figure \ref{evolution_amplitude} shows the temporal evolution of the density amplitude and its position for the traversing and the transmitted waves in two different cases of CH density ($\rho_{CH}=0.1$ marked in blue and $\rho_{CH}=0.3$ marked in red) with regard to a wave having no interaction with a CH (gray). Furthermore, we compare the final density distribution for $\rho_{CH}=0.1$ (blue) and $\rho_{CH}=0.3$ (red) at the end of the simulation run at $t=0.5$. One can see that the waves propagate faster through the CH ($0.4\leq x\leq0.6$) than the primary wave before entering the CH (gray for $0<t<0.2$). Morever, we can observe how the density amplitudes decrease when the wave is traversing through the CH and how they increase again after having left the CH. By comparing the density amplitudes inside the CH, we can see that the amplitude value is much smaller for the case $\rho_{CH}=0.1$ (blue) than in the case $\rho_{CH}=0.3$ (red).

\begin{figure*}[ht]
\centering \includegraphics[width=\textwidth]{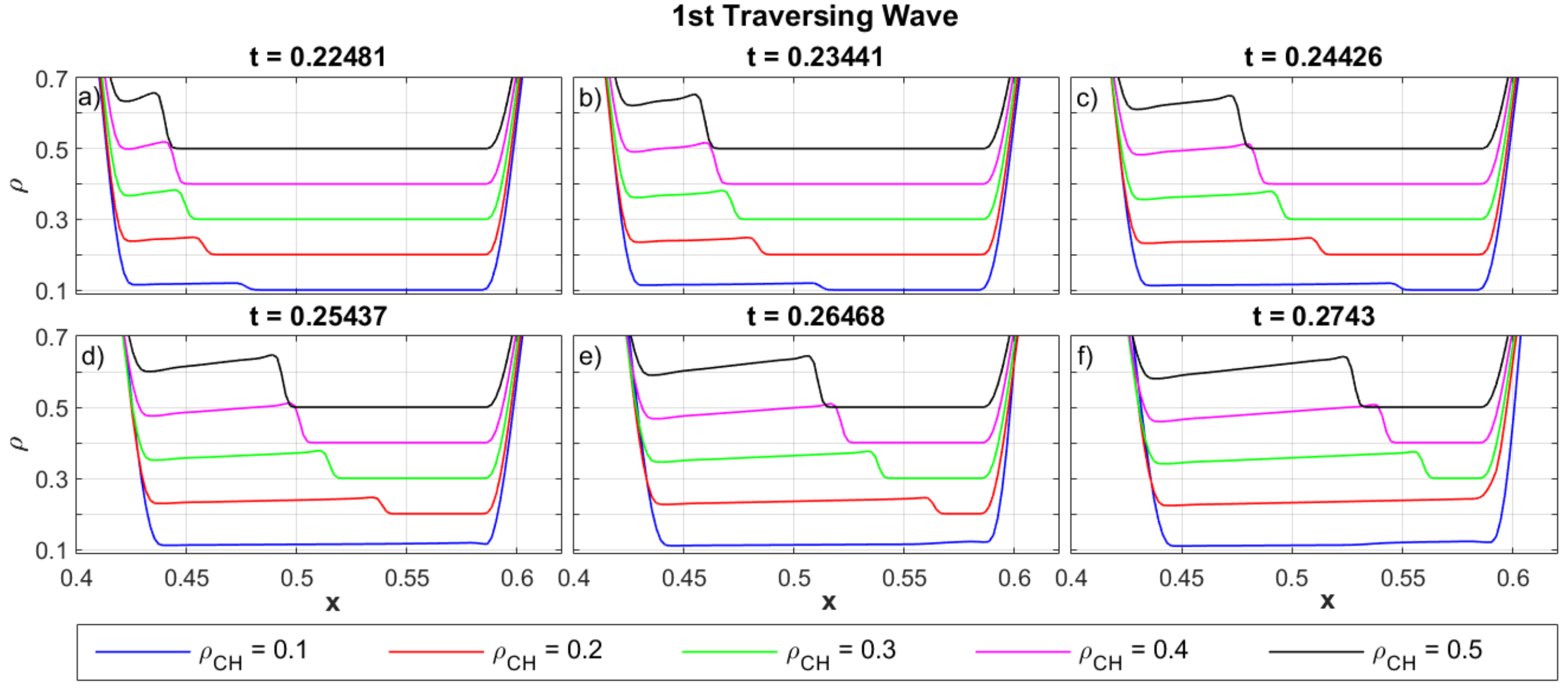}  
\caption{Temporal evolution of the density distribution of the first traversing wave moving in positive $x$-direction inside the CH. Starting shortly after the primary wave has entered the CH ($t=0.22481$) and ending before one part of the wave with the smallest amplitude (blue) gets reflected inside the CH ($t=0.2743$).}
\label{morphology_IN_CH_no0}
\end{figure*}

\begin{figure}[ht]
\centering \includegraphics[width=0.5\textwidth]{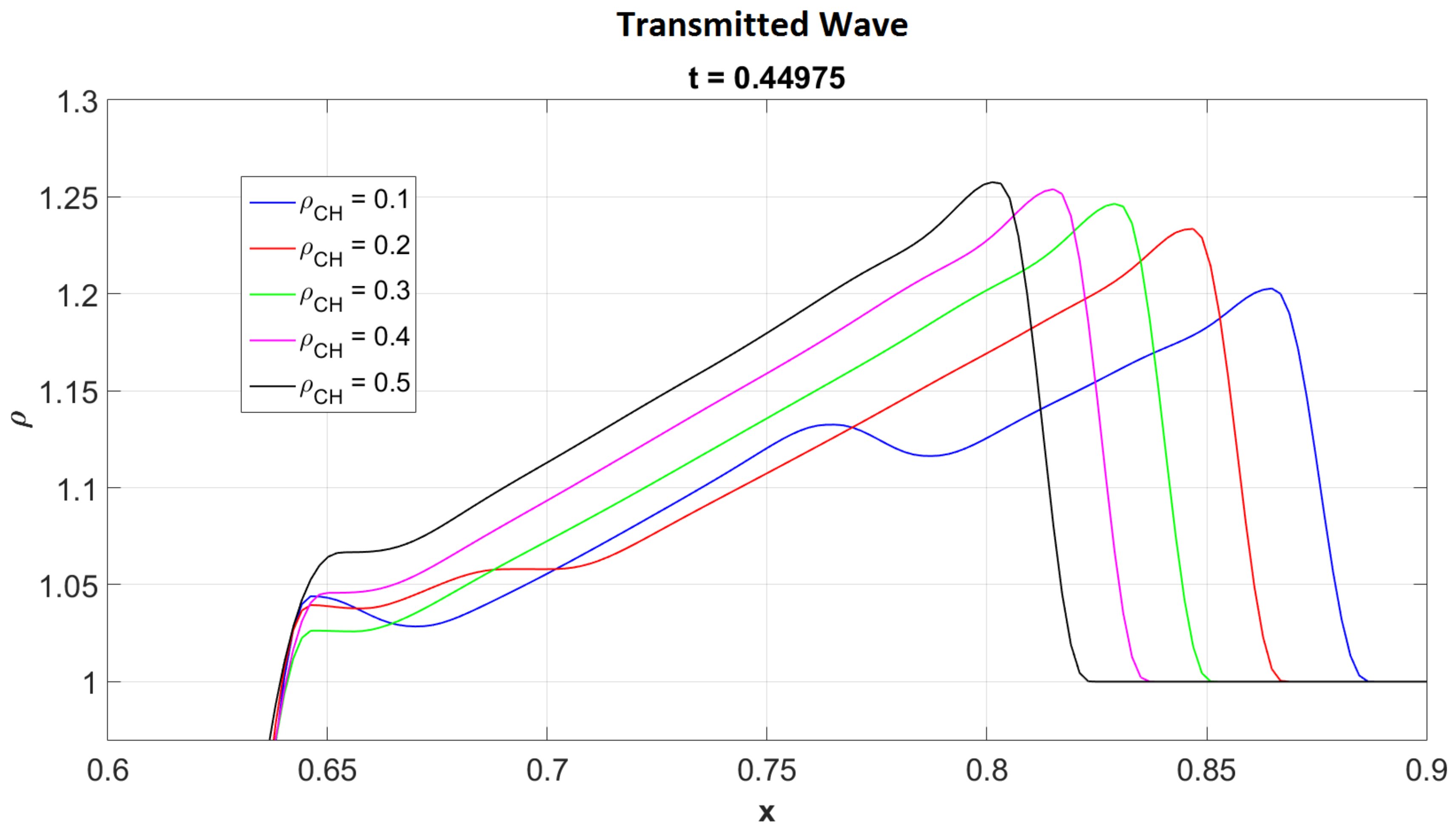}  
\caption{Zoom into the area of the transmitted waves in the range $0.6\leq x\leq0.9$ for the cases $\rho_{CH}=0.1$ (blue), $\rho_{CH}=0.2$ (red), $\rho_{CH}=0.3$ (green), $\rho_{CH}=0.4$ (magenta) and $\rho_{CH}=0.5$ (black). }
\label{zoom_transmitted_wave}
\end{figure}

\begin{figure*}[ht]
\centering \includegraphics[width=\textwidth]{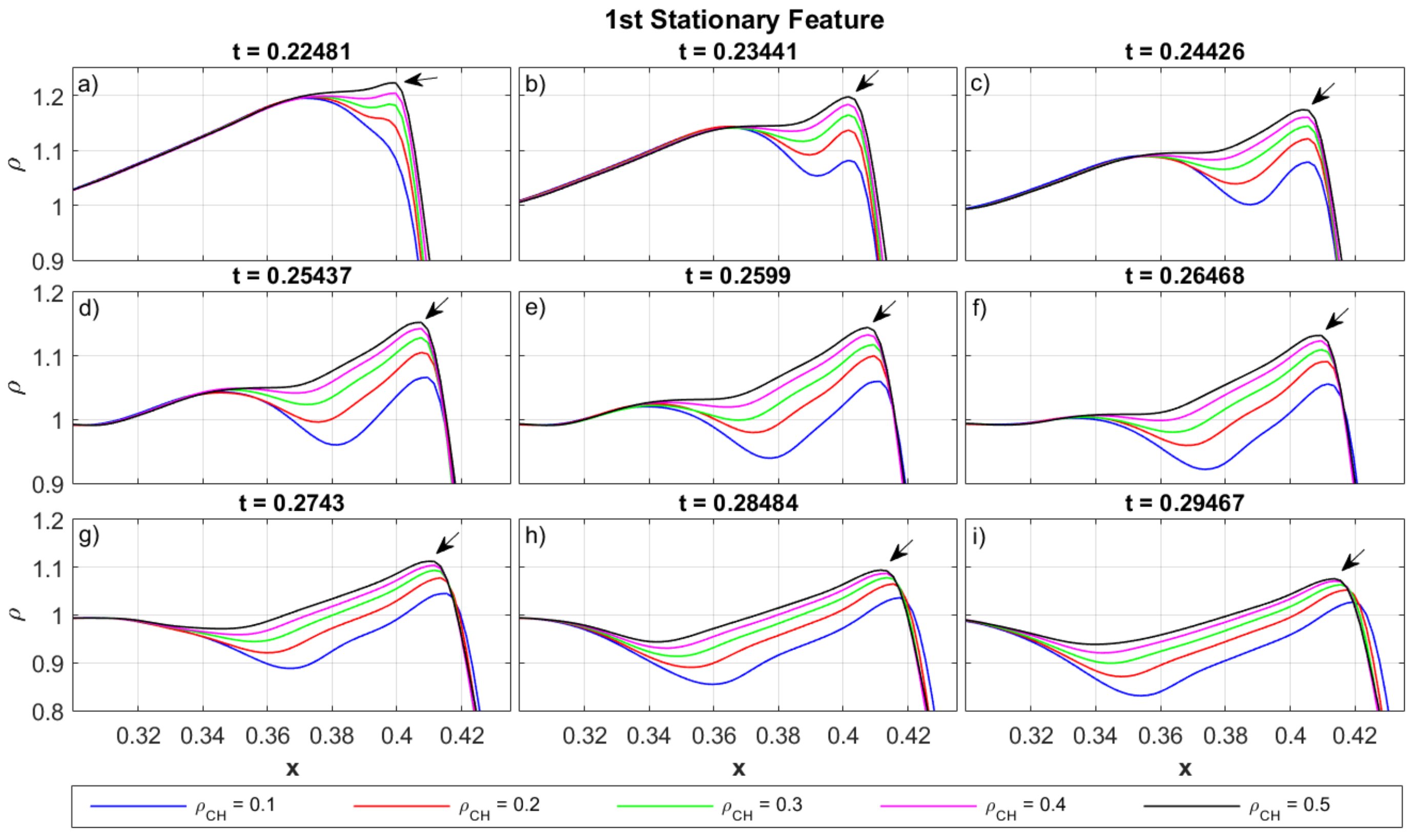}  
\caption{Zoom into the area of the first stationary feature in the range $0.3\leq x\leq0.44$ for the cases $\rho_{CH}=0.1$ (blue), $\rho_{CH}=0.2$ (red), $\rho_{CH}=0.3$ (green), $\rho_{CH}=0.4$ (magenta) and $\rho_{CH}=0.5$ (black). Starting at $t=0.22481$ and ending at $t=0.29467$. The arrows denote the position of the peak value of the first stationary feature for the case $\rho_{CH}=0.5$ (black).}
\label{zoom_first_stat}
\end{figure*}

\begin{figure*}[ht]
\centering \includegraphics[width=\textwidth]{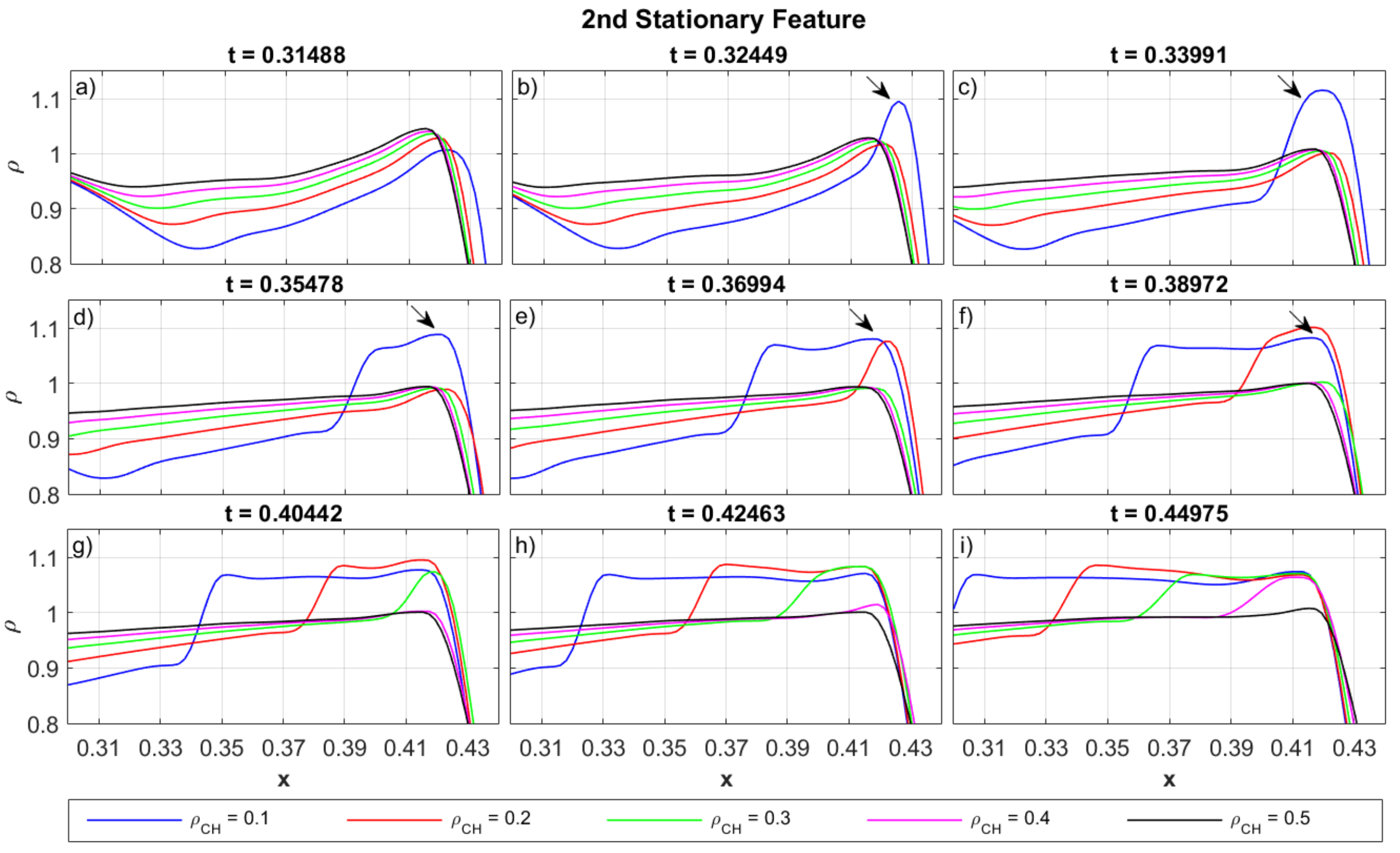}  
\caption{Zoom into the area of the second stationary feature in the range $0.3\leq x\leq0.44$ for the cases $\rho_{CH}=0.1$ (blue), $\rho_{CH}=0.2$ (red), $\rho_{CH}=0.3$ (green), $\rho_{CH}=0.4$ (magenta) and $\rho_{CH}=0.5$ (black). Starting at $t=0.31488$ and ending at $t=0.44975$. The arrows denote the position of the peak value of the second stationary feature for the case $\rho_{CH}=0.1$ (blue).}
\label{zoom_second_stat}
\end{figure*}

\section{Kinematics}

\subsection{Primary Wave}

\begin{figure}[ht!]
\centering\includegraphics[width=0.49\textwidth]{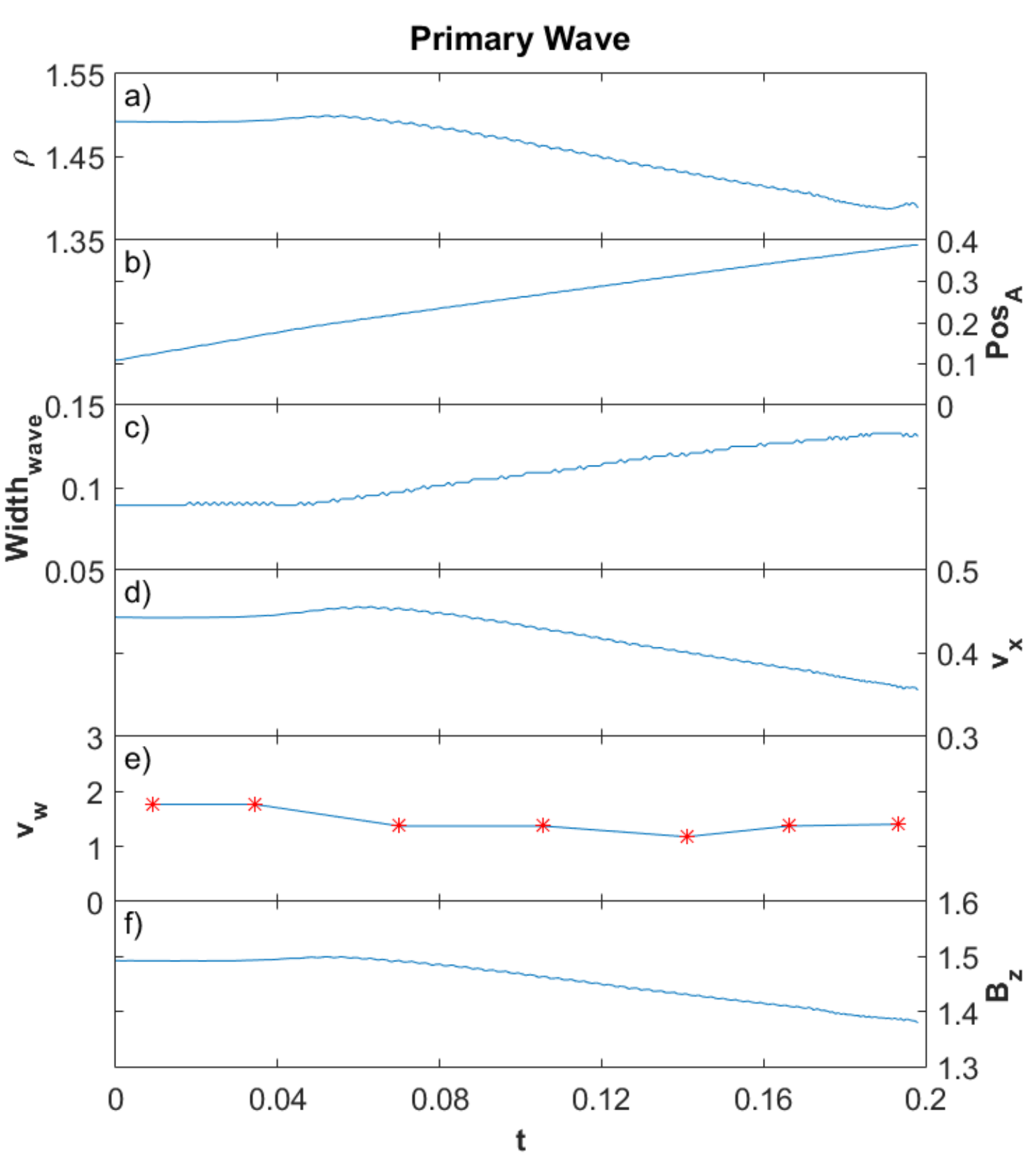}
\caption{From top to bottom: Density, position of the amplitude, width of the wave, plasma flow velocity, phase velocity and magnetic field of the primary wave, from the beginning of the run ($t=0$) until the time when the wave is entering the CH ($t=0.2$). }
\label{Kin_prim_wave}
\end{figure}

Figure \ref{Kin_prim_wave} shows the temporal evolution for the peak values of the primary wave's density, $\rho$, plasma flow velocity, $v_x$, phase speed, $v_w$, and magnetic field in the $z$-direction, $B_z$. In Figure \ref{Kin_prim_wave}a we find that the density amplitude stays approximately constant at a value of $\rho\approx1.5$ until about $t=0.06$ and decreases subsequently to a density value of $\rho\approx1.4$ at $t=0.2$, the time at which the primary wave starts entering the CH. A similar decrease can be seen in Figures \ref{Kin_prim_wave}d and \ref {Kin_prim_wave}f , where we observe the plasma flow velocity and magnetic field component in the $z$-direction decreasing from $v_x\approx0.45$, $B_z\approx1.5$ at $t=0.05$ to $v_x\approx0.35$, $B_z\approx1.4$ at $t=0.2$. At the same time when the amplitude values of $\rho$, $v_x$ and $B_z$ start decreasing, we observe a broadening of the width of the wave, starting at $width_{wave}=0.08$ ($t=0$) and increasing to a value of $width_{wave}=0.13$ ($t=0.2$) (see Figure \ref{Kin_prim_wave}c). Figure \ref{Kin_prim_wave}b shows how the wave is propagating in the positive $x$-direction. In Figure \ref{Kin_prim_wave}e it is evident that the phase speed of the primary wave decreases slightly until the beginning of the entry phase into the CH, i.e. $v_w=1.75$ (at $t=0.01$) decreases to $v_w=1.4$ (at $t=0.2$).

\subsection{Secondary Waves}

Figure \ref{Kin_traver_wave} shows the temporal evolution of density, $\rho$, position of the amplitude, $Pos_{A}$, plasma flow velocity, $v_{x}$, phase speed, $v_{w}$, and magnetic field component in the $z$-direction, $B_{z}$, for the first traversing wave in every case of varying CH density, $\rho_{CH}$. In all five cases the wave is propagating with approximately constant amplitude in the positive $x$-direction (see Figues \ref{Kin_traver_wave}a and \ref{Kin_traver_wave}b), \ie\  $\rho=0.11$ (for $\rho_{CH}=0.1$, blue), $\rho=0.25$ (for $\rho_{CH}=0.2$, red), $\rho=0.39$ (for $\rho_{CH}=0.3$, green), $\rho=0.52$ (for $\rho_{CH}=0.4$, magenta) and $\rho=0.65$ (for $\rho_{CH}=0.5$, black).  Figure \ref{Kin_traver_wave}e shows the values for the magnetic field component in $z$-direction. One can see that, like in the case of the density $\rho$, the amplitudes remain approximately constant and that the smaller the CH density, $\rho_{CH}$, the smaller the amplitude value of $B_{z}$.  The tracking of the wave with $\rho_{CH}=0.5$ starts at a later time due to the different phase speeds of the traversing waves inside the CH. In Figures \ref{Kin_traver_wave}c we observe approximately constant values for $v_x$, but in contrary to the density, $\rho$, and the magnetic field component in $z$-direction, $B_{z}$, the largest amplitudes can be seen in the case of $\rho_{CH}=0.1$ and the smallest ones for $\rho_{CH}=0.5$. The temporal evolution of the phase speed of the first traversing wave is shown in Figure \ref{Kin_traver_wave}d. We find that the smaller the intial density inside the CH, the faster the wave propagates through the CH. In all five cases the phase speed decreases slightly until the wave leaves the CH, \ie\ at $t=0.215$ we have $v_w\approx3.75$ (for $\rho_{CH}=0.1$), $v_w\approx2.5$ (for $\rho_{CH}=0.2$), $v_w\approx2.15$ (for $\rho_{CH}=0.3$) and $v_w\approx1.75$ (for $\rho_{CH}=0.4$). The speed tracking in the case of $\rho_{CH}=0.5$ starts a $t\approx0.222$ and supplies a value of $v_w\approx1.7$. The phase speed values decrease until $t=0.24$ to $v_w\approx3.2$ (for $\rho_{CH}=0.1$), $v_w\approx1.9$ (for $\rho_{CH}=0.2$), $v_w\approx1.9$ (for $\rho_{CH}=0.3$), $v_w\approx1.5$ (for $\rho_{CH}=0.4$) and $v_w\approx1.25$ (for $\rho_{CH}=0.5$). (Due to the very low amplitudes inside the CH one the one hand and the related tracking difficulties on the other hand, there will be no detailed kinematics study of the second and third traversing wave.)

The temporal evolution of the parameters of the transmitted waves is described in Figure \ref{Kin_transm_wave}. In Figures \ref{Kin_transm_wave}a, \ref{Kin_transm_wave}b and \ref{Kin_transm_wave}e, where one can see the amplitude values of $\rho$, $v_x$ and $B_z$, it is evident that the wave which was traversing through the CH in the case $\rho_{CH}=0.1$ (blue line) leaves the CH first, followed by the waves in the cases of $\rho_{CH}=0.2$ (red), $\rho_{CH}=0.3$ (green), $\rho_{CH}=0.4$ (magenta) and $\rho_{CH}=0.5$ (black). The density amplitude values of the transmitted waves start at $\rho=1.27$ (for $\rho_{CH}=0.1$ at $t\approx0.28$), $\rho=1.3$ (for $\rho_{CH}=0.2$ at $t\approx0.3$), $\rho=1.32$ (for $\rho_{CH}=0.3$ at $t\approx0.32$), $\rho=1.325$ (for $\rho_{CH}=0.4$ at $t\approx0.33$) and $\rho=1.325$ (for $\rho_{CH}=0.5$ at $t\approx0.337$) and decrease to $\rho=1.24$ (for $\rho_{CH}=0.1$), $\rho=1.28$ (for $\rho_{CH}=0.2$), $\rho=1.29$ (for $\rho_{CH}=0.3$), $\rho=1.3$ (for $\rho_{CH}=0.4$) and $\rho=1.305$ (for $\rho_{CH}=0.5$) at the end of the simulation run at $t=0.5$ (see Figure \ref{Kin_transm_wave}a).  Figure \ref{Kin_transm_wave}c shows how the transmitted waves propagate in the positive $x$-direction in all five cases. The evolution of the phase speed of the transmitted waves is described in Figure \ref{Kin_transm_wave}d. Here we can see that the values start at  $v_w=1.21$ (for $\rho_{CH}=0.1$), $v_w=1.18$ (for $\rho_{CH}=0.2$), $v_w=1.17$ (for $\rho_{CH}=0.3$), $v_w=1.17$ (for $\rho_{CH}=0.4$) and $v_w=1.17$ (for $\rho_{CH}=0.5$) and decrease slightly in all five cases as the wave is moving further towards the positive $x$-direction.

Figure \ref{Kin_first_reflection} describes the amplitude values of the first reflection. Due to a superposition, caused by a simultaneous entering of segments of the rear of the primary wave into the CH on the one hand and an already ongoing reflection of the the front segments of the wave on the other hand, this feature is not able to move in the negative $x$-direction until the primary wave has completed its entry phase into the CH. Hence, we will start describing the kinematics of this first reflection at $t\approx0.27$, when it starts moving in the negative $x$-direction.

As we can see in Figures \ref{morphology_1D_part2}a - \ref{morphology_1D_part2}e, the first reflection is the same in all cases of different $\rho_{CH}$.  It moves from $x\approx0.3$ (seen in Figure \ref{morphology_1D_part2}a) to $x\approx0.2$ (seen in Figure \ref{morphology_1D_part2}d) and is located at the left side of the density depletions.  Figure \ref{Kin_first_reflection} describes the kinematics of this first reflection for all different $\rho_{CH}$. In Figure \ref{Kin_first_reflection}a one can see that the amplitude density stays at an approximately constant value of about $\rho=1.0$ until $t\approx0.39$. At that time the first reflection approaches an area of oscillations that is caused by numerical effects (detailed description in \citealt{Piantschitsch2017}). Here we can no longer get reasonable results for the first reflection. Similar to the density values of this first reflection the magnetic field component, $B_{z}$, and plasma flow velocity, $v_{x}$, stay approximately constant at values of $B_{z}=0.495$ or $v_{x}=0.001$  (see Figure \ref{Kin_first_reflection}c and Figure \ref{Kin_first_reflection}e). Figure \ref{Kin_first_reflection}b shows how the reflection is moving in the negative $x$-direction. The temporal evolution of the phase speed of this first reflection is described in Figure \ref{Kin_first_reflection}d. Here we observe that the value of the phase speed decreases from $v_{w}\approx-1.1$ to $v_{w}\approx-0.5$.

In Figure \ref{Kin_second_reflection} we present the kinematic analysis of the second reflection. This reflection is caused by parts of the traversing wave leaving the CH at the left CH boundary at $t\approx0.36$. At the time we stop the simulation run at $t=0.5$ only the reflections for the cases $\rho_{CH}=0.1$, $\rho_{CH}=0.2$ and $\rho_{CH}=0.3$ have moved sufficiently far in the negative $x$-direction to compare their peak values. Figure \ref{Kin_second_reflection} shows how the time at which the second reflection appears depends on the density inside the CH. In contrast to traversing and transmitted waves we do not have a linear correlation between the initial density values inside the CH and the amplitude values of the different reflection parameters. Figure \ref{Kin_second_reflection}a shows that the density amplitude for the case $\rho_{CH}=0.2$ (red) is in fact larger than the density amplitude in the case $\rho_{CH}=0.1$ (blue). However, the density amplitude in the case $\rho_{CH}=0.3$ (green) is not larger than the one in the case $\rho_{CH}=0.2$ (red) but lies between the first two cases. A similar behaviour holds true for the plasma flow velocity, $v_x$, and the magnetic field component, $B_z$ (see Figures \ref{Kin_second_reflection}b and \ref{Kin_second_reflection}d). In Figure \ref{Kin_second_reflection}c one can see how the second reflection is moving in the negative $x$-direction until the end of the simulation run at $t=0.5$.

\subsection{Stationary Features}

\begin{figure}[ht!]
\centering\includegraphics[width=0.49\textwidth]{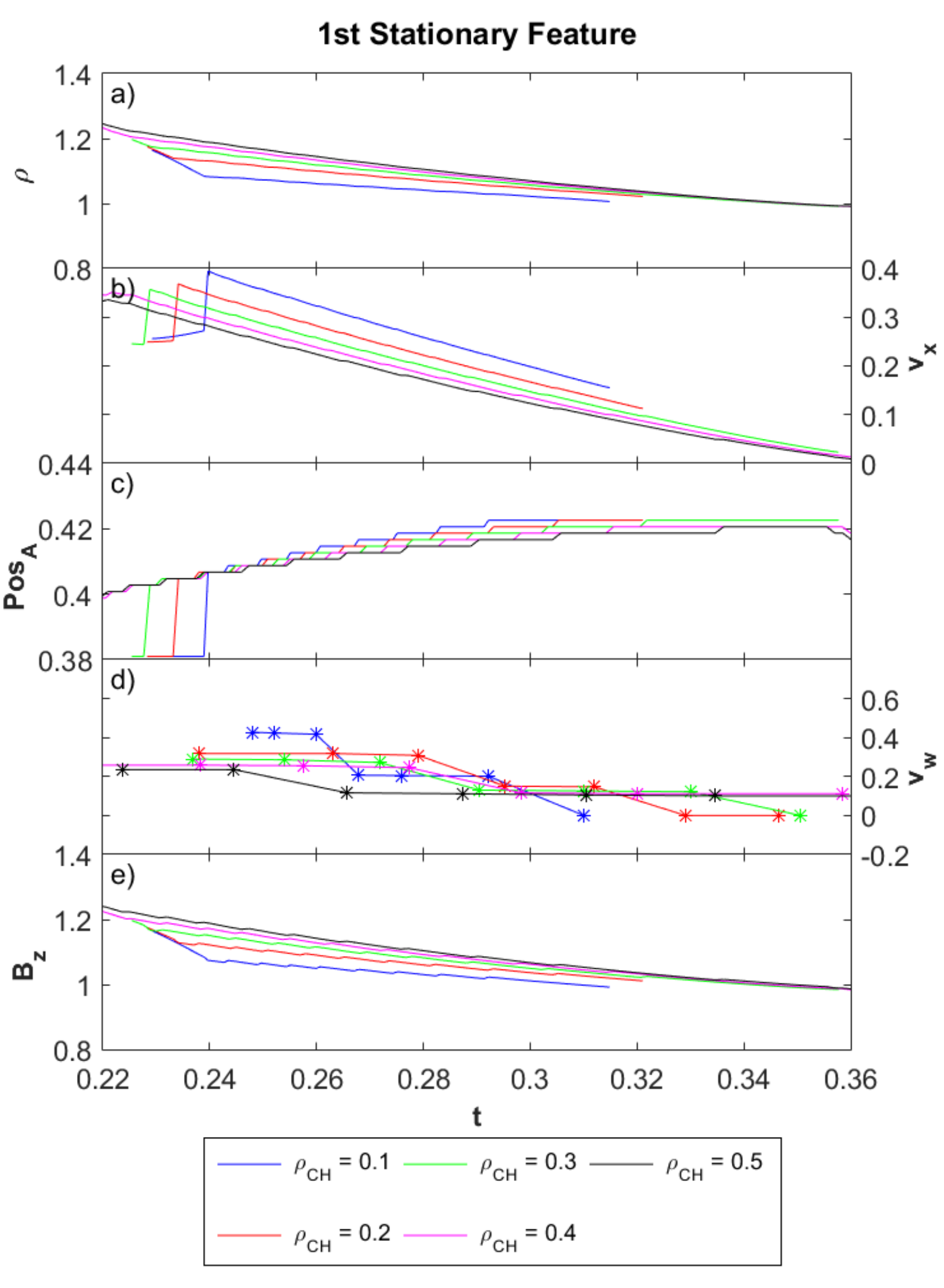}
\caption{From top to bottom: Temporal evolution of density, plasma flow velocity, position of the amplitude, phase velocity and magnetic field of the first stationary feature for the cases $\rho_{CH}=0.1$ (blue), $\rho_{CH}=0.2$ (red), $\rho_{CH}=0.3$ (green), $\rho_{CH}=0.4$ (magenta) and $\rho_{CH}=0.5$ (black). Starting at about $t=0.22$, when this feature occurs first in case of $\rho_{CH}=0.5$ (black) and ending at $t\approx0.36$.}
\label{Kin_first_stat}
\end{figure}

The kinematics of the first stationary feature are described in Figure \ref{Kin_first_stat}. In Figure \ref{Kin_first_stat}a we can see that at about $t\approx0.22$ this feature occurs first in the case of $\rho_{CH}=0.5$ (black), starting with a density amplitude of $\rho\approx1.25$ and decreasing to $\rho\approx1.0$ at $t=0.36$. This density plot also shows that the appearance of the other density amplitudes follow one after each other: $\rho=1.23$ (for $\rho_{CH}=0.4$, magenta, at $t=0.22$), $\rho=1.2$ (for $\rho_{CH}=0.3$, green, at $t=0.225$), $\rho=1.18$ (for $\rho_{CH}=0.2$, red, at $t=0.23$) and $\rho=1.16$ (for $\rho_{CH}=0.1$, blue at $t=0.229$). Finally, these amplitude values decrease to $\rho\approx1.0$ (in all five cases of different $\rho_{CH}$).
A similar decreasing behaviour can be observed for the magnetic field component in the $z$-direction, $B_z$, (see Figure \ref{Kin_first_stat}e). The amplitude values of $B_z$ start at approximately the same values as the density amplitudes in Figure \ref{Kin_first_stat}a and decrease also to a value of $B_{z}=1.0$ at $t=0.36$. Figures \ref{Kin_first_stat}a and \ref{Kin_first_stat}e show that the smaller the density inside the CH, the larger the  wave's amplitude values for density and magnetic field. The exact reverse behaviour can be observed for the plasma flow velocity, $v_{x}$, and phase speed, $v_{w}$, of this feature (see Figures \ref{Kin_first_stat}b and \ref{Kin_first_stat}d). Here it is evident that the smaller the density value inside the CH, the smaller the values for $v_{x}$ and $v_{w}$. Figure \ref{Kin_first_stat}c shows that the first stationary feature is moving slightly in the positive $x$-direction in all five cases of different $\rho_{CH}$.

\begin{figure}[ht!]
\centering\includegraphics[width=0.49\textwidth]{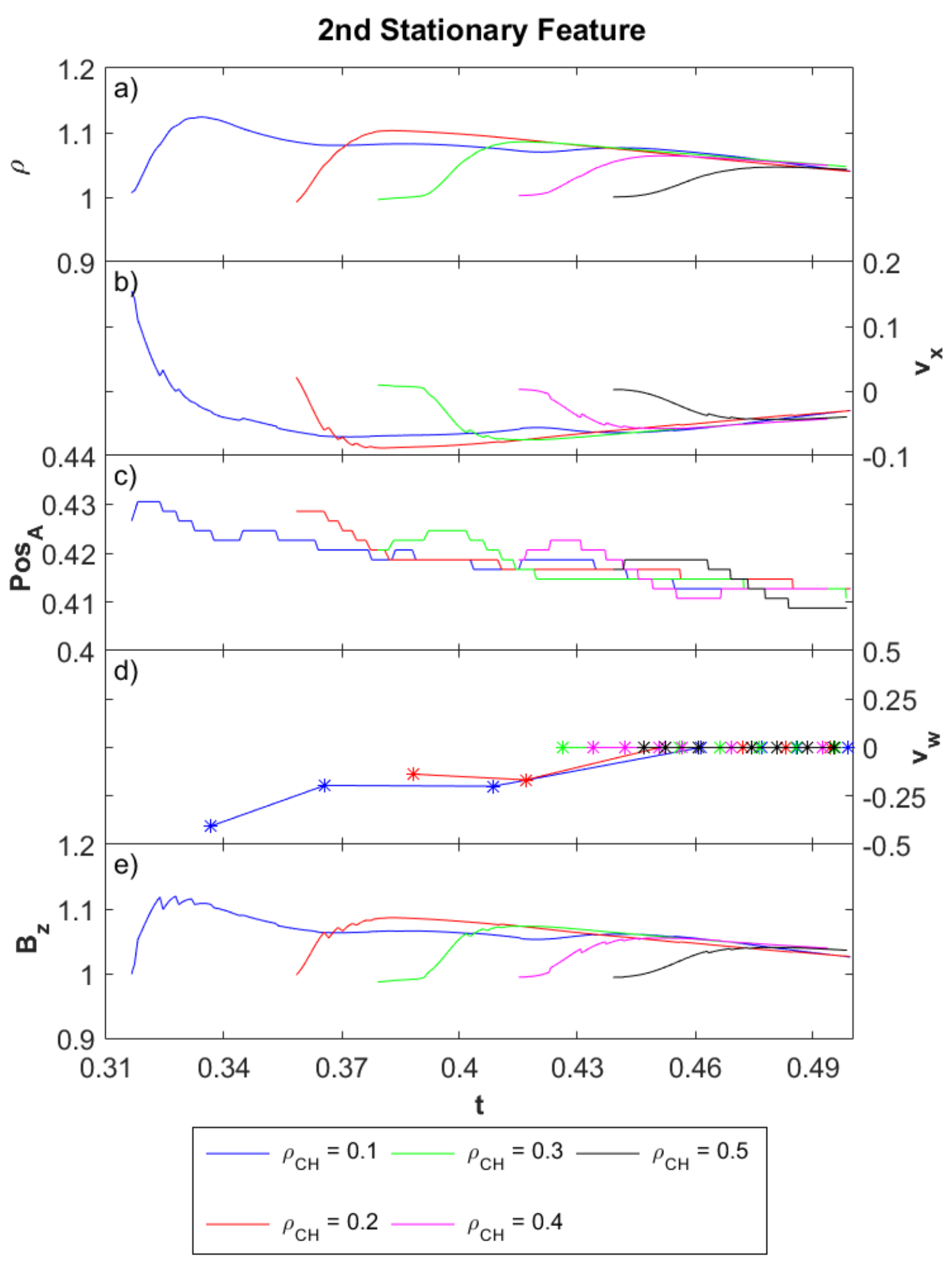}
\caption{From top to bottom: Temporal evolution of density, plasma flow velocity, position of the amplitude, phase velocity and magnetic field of the second stationary feature for the cases $\rho_{CH}=0.1$ (blue), $\rho_{CH}=0.2$ (red), $\rho_{CH}=0.3$ (green), $\rho_{CH}=0.4$ (magenta) and $\rho_{CH}=0.5$ (black). Starting at about $t=0.32$, when this feature occurs first in case of $\rho_{CH}=0.1$ (blue) and ending at the end of the simulation run at $t=0.5$.}
\label{Kin_second_stat}
\end{figure}

In Figure \ref{Kin_second_stat} we present the kinematic analysis of the second stationary feature. Figures \ref{Kin_second_stat}a and \ref{Kin_second_stat}e show, in contrast to the first stationary feature, that the smaller the initial density inside the CH, $\rho_{CH}$, the larger the amplitude values for density, $\rho$, and magnetic field component, $B_z$. Another difference between the first and the second stationary features is the fact, that the second stationary feature is moving slightly in the negative $x$-direction (see Figure \ref{Kin_second_stat}c). Since the movement of this feature is more or less only a small shift of its position to the left, plasma flow velocity $v_x$ and phase speed $v_w$ are very small too and finally decrease to a value of almost zero in all five cases of different $\rho_{CH}$ (see Figures \ref{Kin_second_stat}b and \ref{Kin_second_stat}d).

\subsection{Density Depletion}

\begin{figure}[ht!]
\centering\includegraphics[width=0.49\textwidth]{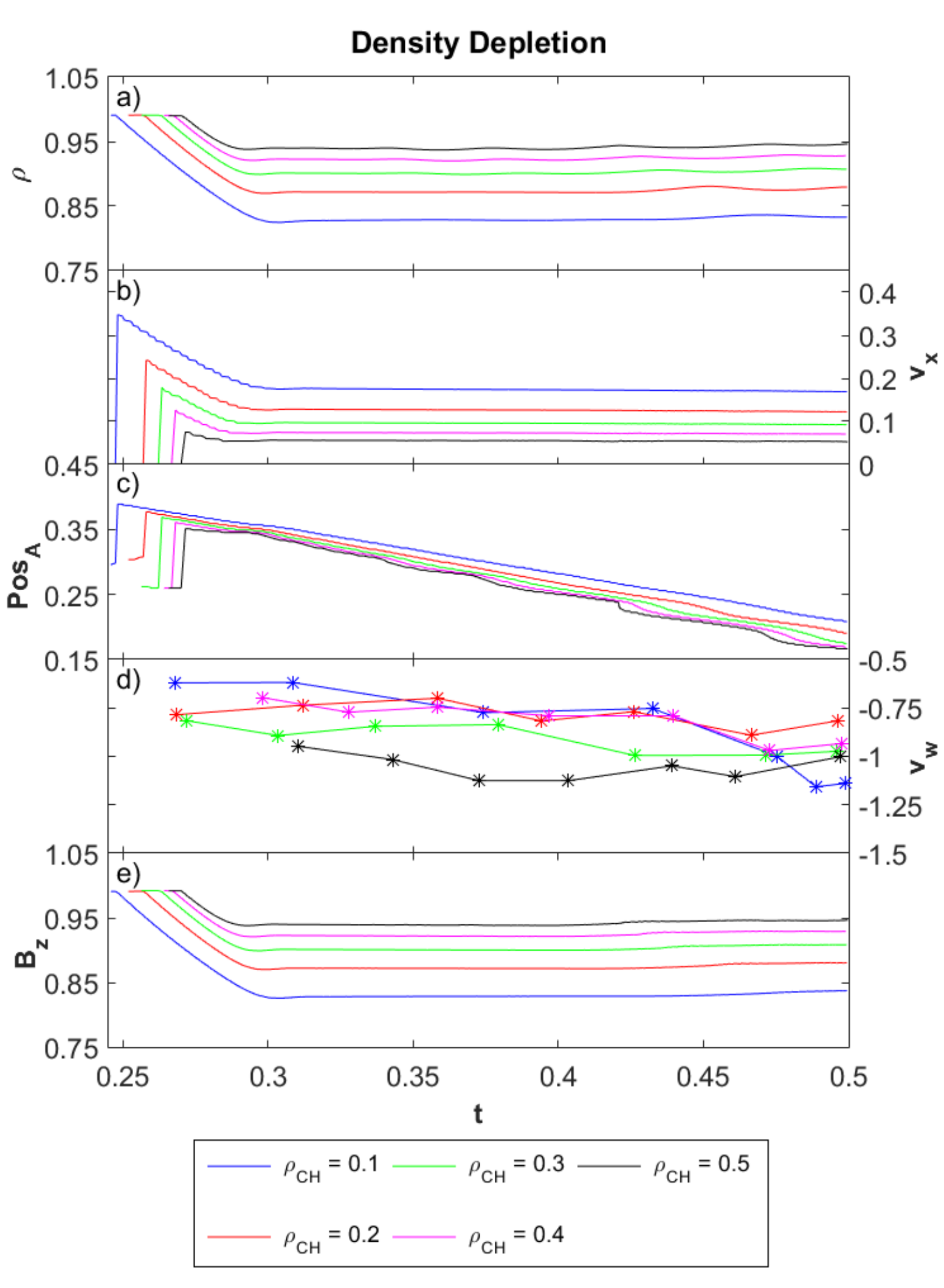}
\caption{From top to bottom: Temporal evolution of density, plasma flow velocity, position of the amplitude, phase velocity and magnetic field of the density depletion for the cases $\rho_{CH}=0.1$ (blue), $\rho_{CH}=0.2$ (red), $\rho_{CH}=0.3$ (green), $\rho_{CH}=0.4$ (magenta) and $\rho_{CH}=0.5$ (black). Starting at about $t=0.25$, when this feature occurs first in case of $\rho_{CH}=0.1$ (blue) and ending at the end of the simulation run at $t=0.5$.}
\label{Kin_density_depletion}
\end{figure}

In Figure \ref{Kin_density_depletion} we analyze the temporal evolution of the density depletion for all different cases of initial CH density. This feature occurs first for the case $\rho_{CH}=0.1$ (blue) at about $t=0.25$, followed by the density depletions for $\rho_{CH}=0.2$ (red), $\rho_{CH}=0.3$ (green), $\rho_{CH}=0.4$ (magenta) and $\rho_{CH}=0.5$ (black). The minimum density values of the depletion decrease from about $\rho=1.0$ (for all five cases) to $\rho=0.83.$ ($\rho_{CH}=0.1$, blue), $\rho=0.87$ ($\rho_{CH}=0.2$, red), $\rho=0.9$ ($\rho_{CH}=0.3$, green), $\rho=0.92$ ($\rho_{CH}=0.4$, magenta) and $\rho=0.94$ ($\rho_{CH}=0.5$, black) at $t=0.3$ and subsequently remain approximately constant at those values until the end of the run at $t=0.5$ (see Figure \ref{Kin_density_depletion}a). An analogous behaviour to the density evolution can be found for the temporal evolution of the magnetic field component $B_{z}$ (see Figure \ref{Kin_density_depletion}e). In Figure \ref{Kin_density_depletion}b we find a decrease of the plasma flow velocity for all five cases of different intial CH density $\rho_{CH}$. At $t=0.25$ we find a value of $v_{x}\approx0.35$ which decreases down to $v_{x}\approx0.17$ (at $t=0.3$) in the case of $\rho_{CH}=0.1$ (blue). A similar decrease can be found for all the other cases: The density values $v_{x}\approx0.24$ (red, at $t\approx0.258$), $v_{x}\approx0.18$ (green, at $t\approx0.26$), $v_{x}\approx0.13$ (magenta, at $t\approx0.268$) and $v_{x}\approx.0.08$ (black, at $t\approx0.27$) decrease to $v_{x}\approx0.13.$ (red), $v_{x}\approx0.1$ (green), $v_{x}\approx0.07$ (magenta) and $v_{x}\approx0.05$ (black) and then remain at those values until $t=0.5$. Figure \ref{Kin_density_depletion}c shows how all density depletions are moving towards the negative $x$-direction. Furthermore, we observe that the smaller the density inside the CH, the smaller the mean phase speed of the density depletion. (see Figure \ref{Kin_density_depletion}d).

\begin{figure}[ht!]
\centering\includegraphics[width=0.49\textwidth]{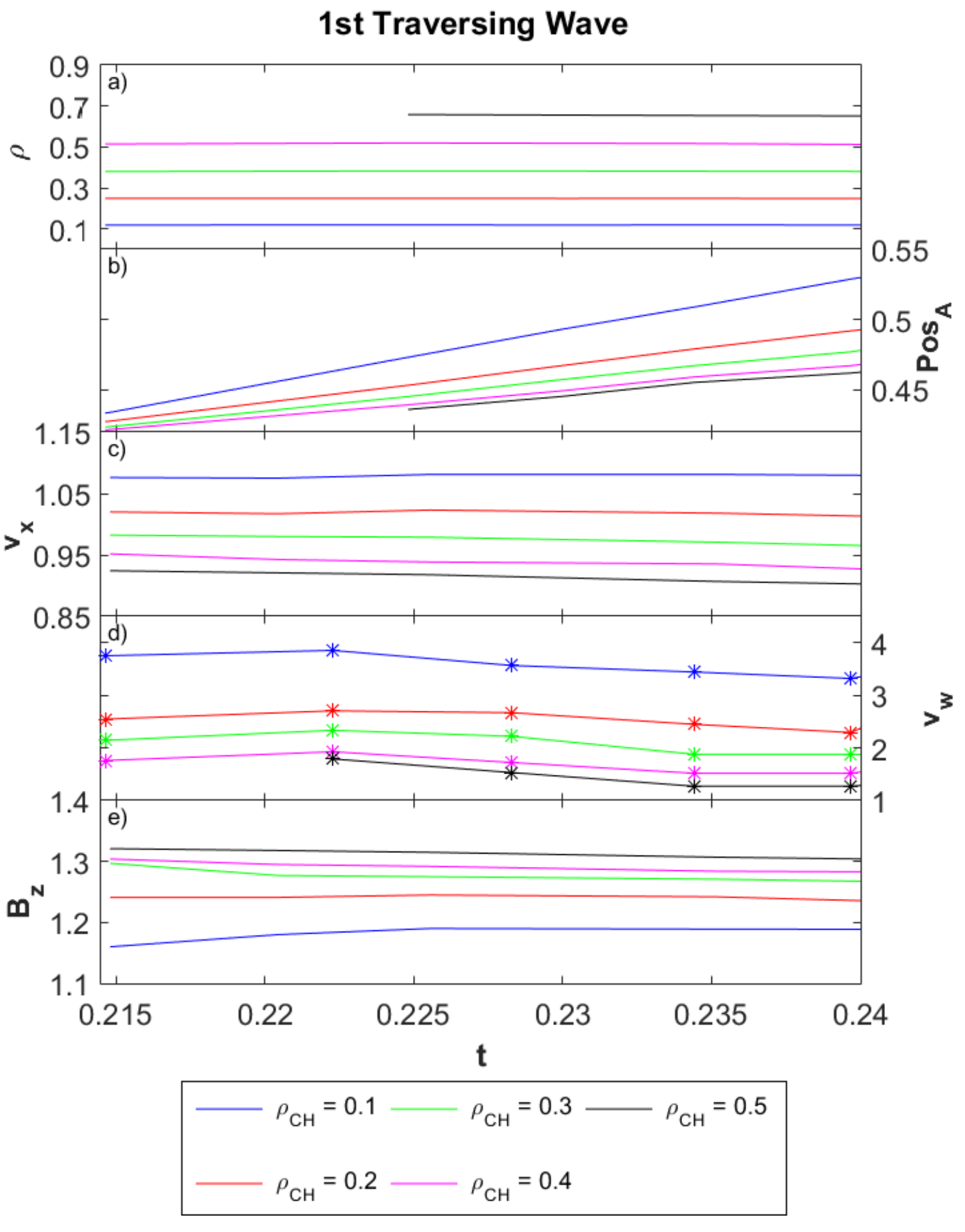}
\caption{From top to bottom: Temporal evolution of density, position of the amplitude, plasma flow velocity, phase velocity and magnetic field of the traversing wave for the cases $\rho_{CH}=0.1$ (blue), $\rho_{CH}=0.2$ (red), $\rho_{CH}=0.3$ (green), $\rho_{CH}=0.4$ (magenta) and $\rho_{CH}=0.5$ (black). Starting at about $t=0.215$, shortly after the primary wave has entered the CH and ending at $t=0.24$, when the traversing waves leave the CH.}
\label{Kin_traver_wave}
\end{figure}

\begin{figure}[ht!]
\centering\includegraphics[width=0.49\textwidth]{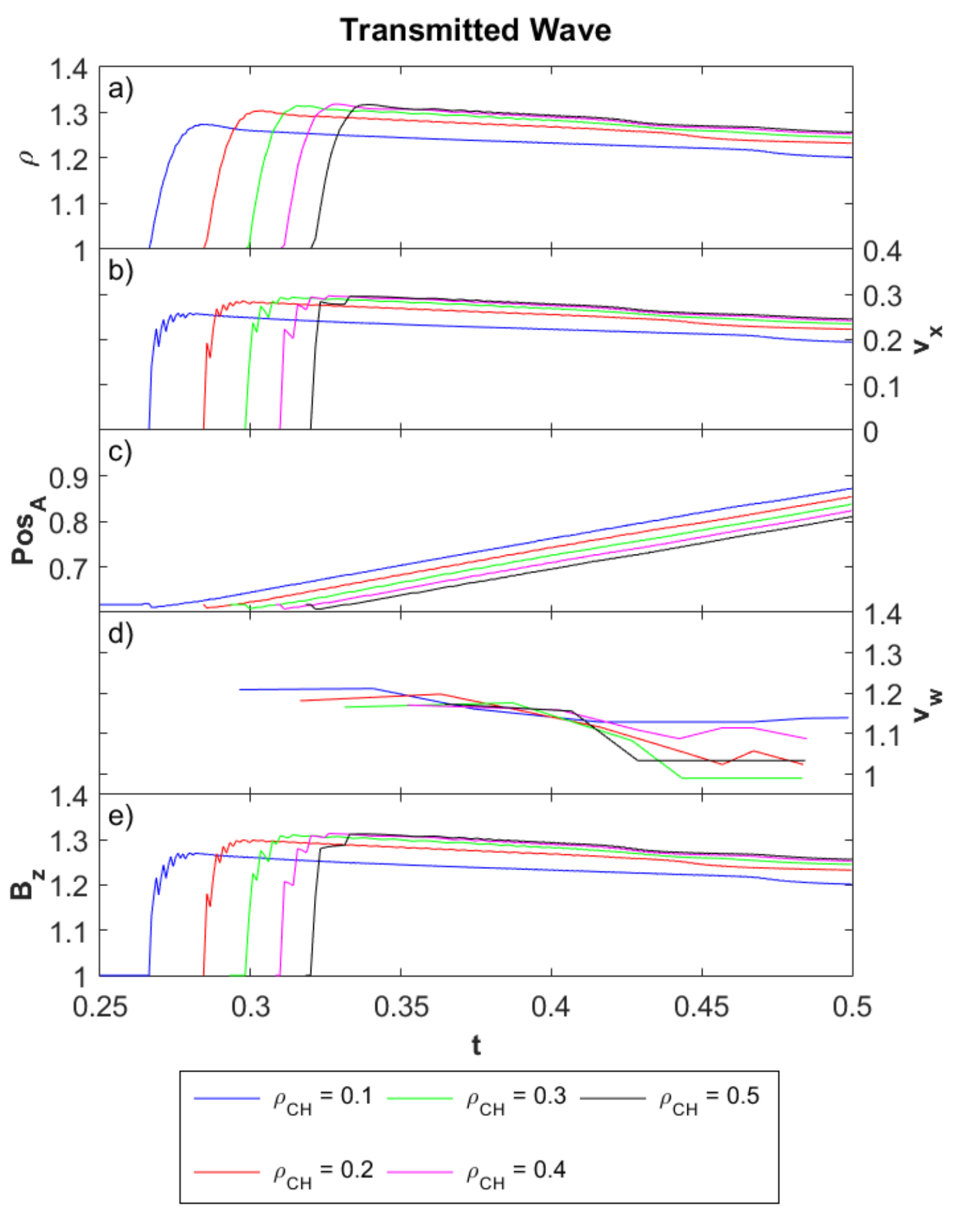}
\caption{From top to bottom: Temporal evolution of density, plasma flow velocity, position of the amplitude, phase velocity and magnetic field of the transmitted wave for the cases $\rho_{CH}=0.1$ (blue), $\rho_{CH}=0.2$ (red), $\rho_{CH}=0.3$ (green), $\rho_{CH}=0.4$ (magenta) and $\rho_{CH}=0.5$ (black). Starting at about $t=0.26$, when the first transmitted wave (blue) occurs at the right CH boundary and ending at the end of the simulation run at $t=0.5$.}
\label{Kin_transm_wave}
\end{figure}

\begin{figure}[ht!]
\centering\includegraphics[width=0.49\textwidth]{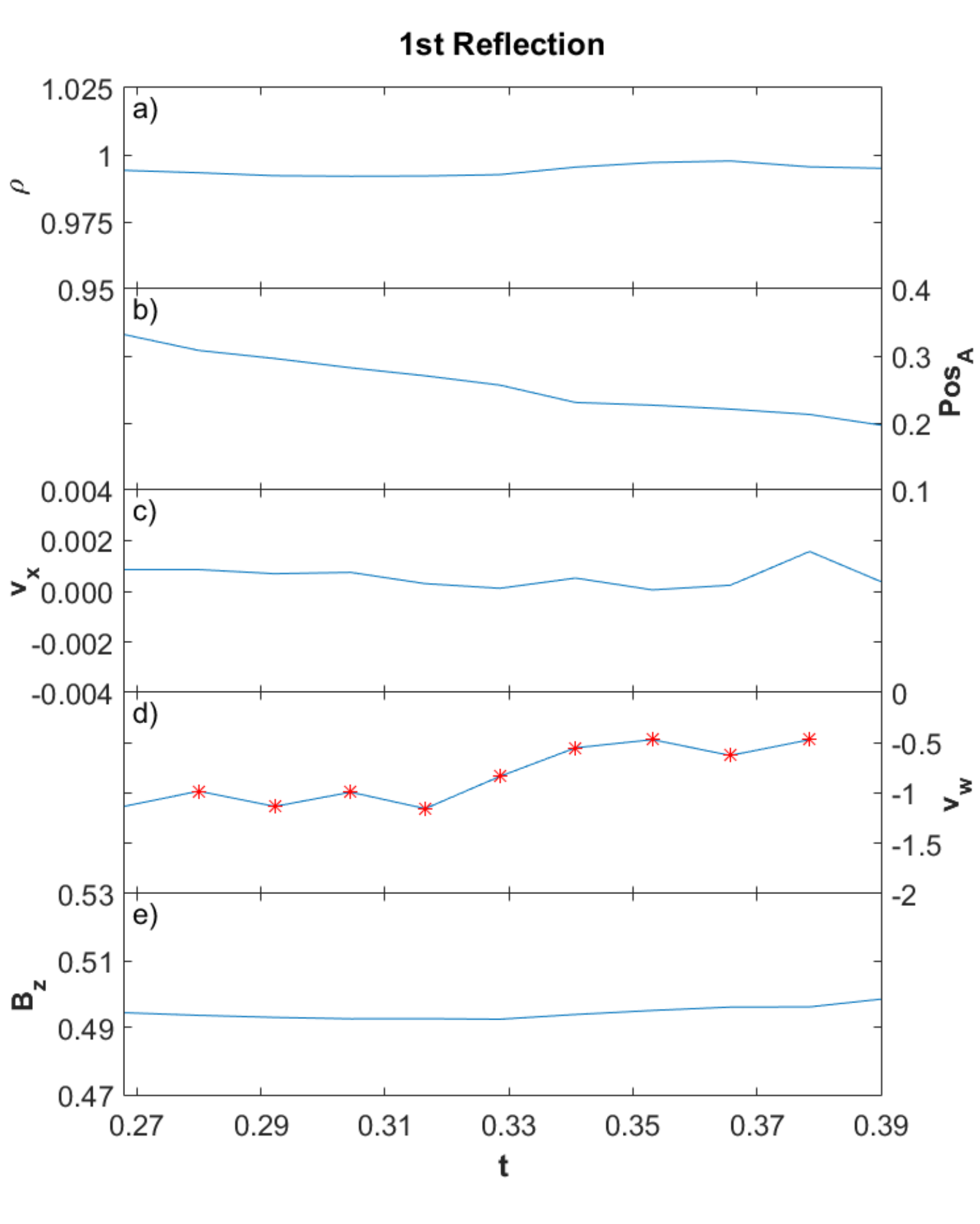}
\caption{From top to bottom: Temporal evolution of density, position of the amplitude, plasma flow velocity, phase velocity and magnetic field of the first reflection starting at ($t\approx0.27$) and ending at ($t=0.39$).}
\label{Kin_first_reflection}
\end{figure}

\begin{figure}[ht!]
\centering\includegraphics[width=0.49\textwidth]{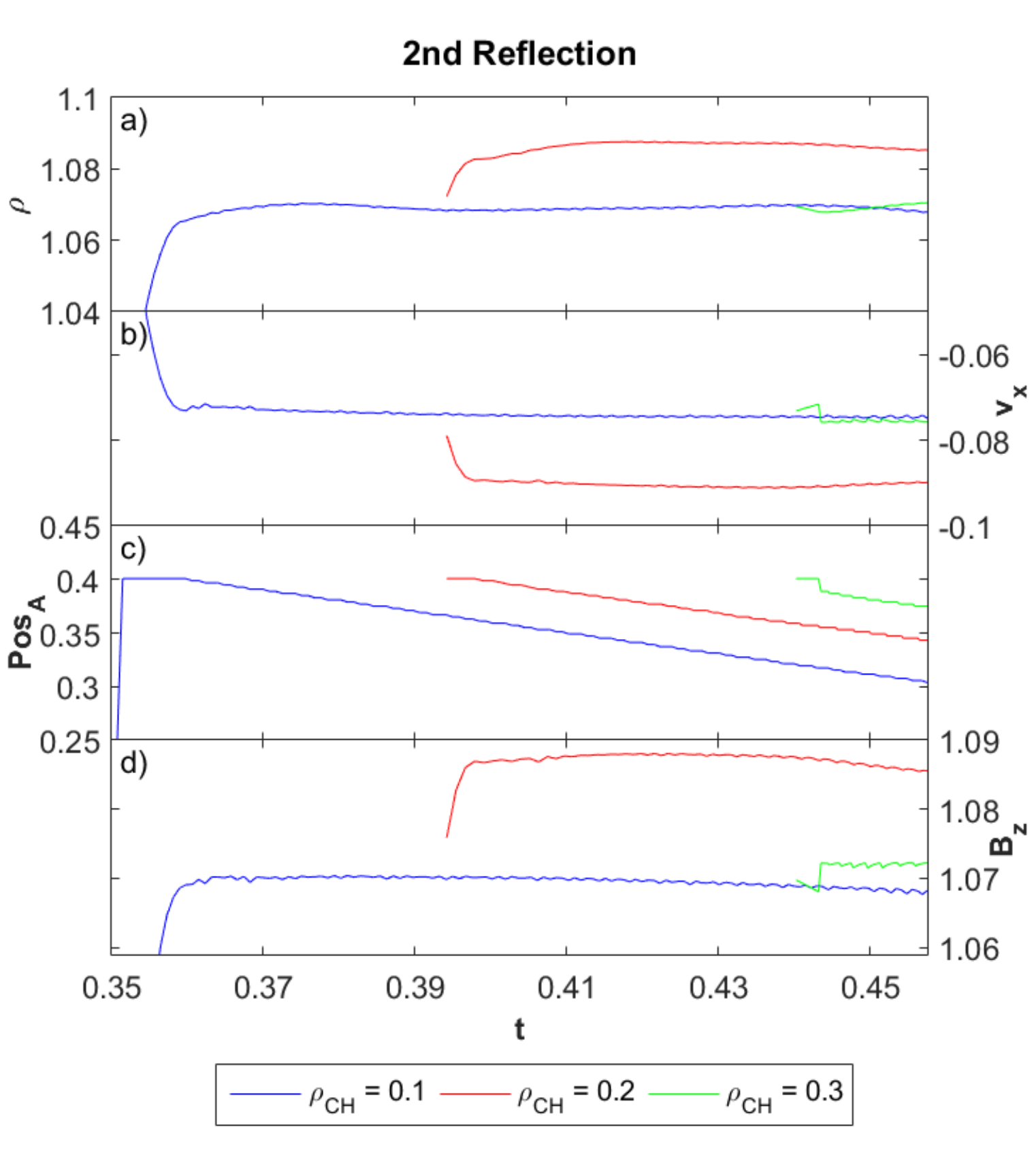}
\caption{From top to bottom: Temporal evolution of density, plasma flow velocity, position of the amplitude and magnetic field of the second reflection for the cases $\rho_{CH}=0.1$ (blue), $\rho_{CH}=0.2$ (red) and $\rho_{CH}=0.3$ (green). Starting at about $t=0.35$, when the second reflection occurs for the case $\rho_{CH}=0.1$ (blue) and ending at the end of the simulation run at $t=0.5$.}
\label{Kin_second_reflection}
\end{figure}

\section{Discussion}

In \citet{Piantschitsch2017} we showed that the impact of the incoming fast-mode MHD wave on the CH leads to effects like reflection, transmission and the formation of stationary fronts. In this paper we focus on how the CH density influences all these different features. We find that the CH density correlates with the peak values of the stationary features and the amplitudes of the secondary waves.

When we compare the first reflection and the first stationary feature with each other, we see that both effects are connected to a superposition of wave parts which are entering the CH and wave parts that were already reflected at the CH boundary. In detail this means, that segments of the rear of the primary wave are entering the CH while segments of the front of the primary wave have already been reflected at the left CH boundary but are prevented from moving in the negative $x$-direction due to the plasma flow associated with the primary wave. The significant difference between those two features is, that the first reflection is a moving feature and its parameters are the same for all five cases of different CH density. The first stationary feature on the other hand exhibits different amplitudes depending on the various CH densities. This means, further, that the first reflection is only caused by the immediate response of the primary's wave impact on the CH boundary. The first stationary feature in contrary seems to be also affected by the different CH densities. 

A comparison between the first and the second stationary feature shows that these effects depend on the initial CH density in the opposite manner. The amplitudes for the second stationary feature are larger, the smaller the initial CH density is. The density amplitudes of the first stationary feature in contrary are smaller, the smaller the initial CH density is. An explanation for this could be a combination of the effects of the traversing waves on the one hand and reflections inside the CH on the other hand in the case of the second stationary feature. We showed in the kinematics section that the smaller the initial CH density, the smaller the density amplitude of the traversing wave and the smaller the density amplitude of the transmitted wave. This consequently also means that in the case of low initial CH density a bigger part gets reflected inside the CH and leads finally to a larger peak value of the second stationary feature. 

The second reflection exhibits notable properties since there is no linear correlation between its amplitudes and the initial CH density. This feature seems to be more complex since it combines effects of the traversing waves, their phase speed and reflections inside the CH. 

During the analysis of the transmitted waves, we found an additional peak inside the wave, a kind of subwave which is moving with the transmitted wave in the positive $x$-direction. This phenomenon only occurs for the cases of $\rho_{CH}=0.1$ and $\rho_{CH}=0.2$. A reason for this is probably the limited runtime of the simulation, \ie\ we expect to see those peaks in the transmitted waves for the other three cases as well for a longer runtime of the simulation. We found that these peaks occur when the third traversing wave reaches the right CH boundary.

When considering our simulation results we have to bear in mind that we are dealing with an idealized situation including many constraints, \eg\ a homogeneous magnetic field, the fact that the pressure is equal to zero over the whole computational box, the assumption of a certain value for the initial wave amplitude and a simplified shape of the CH. Another thing we have to pay attention to is the fact that in our simulations we assume a certain width of the CH. We do not know so far how much a broader CH would influence the final phase speed of the traversing waves and hence the properties of the transmitted waves as well as the reflective features inside the CH.

In our simulations we observe a quite large density amplitude of the transmitted wave whereas in observations such transmitted waves are rarely found. Only in \citet{Olmedo2012} for the first time a wave reported being transmitted through a CH. Hence, there are some aspects, like the intensity of the wave's driver (solar flare or CME), the distance of the initial wave front to the CH, the shape and size of the CH and the magnetic field structure inside the CH, that we have to keep in mind when comparing observations with our simulations. More specifically, in our simulations we make sure that the amplitude of the incoming wave is large enough and that the distance to the CH is sufficiently small in order to guarantee a transmission through the CH. In observations, due to a possibly weak eruption or a large distance of the wave's driver to the CH, this can not be guaranteed. Another issue is the shape of CH; in our simulations the wave is approaching exactely perpendicular to the CH at every point, whereas this is usually not the case in the observations. Moreover, the size of a CH can also be a reason for preventing a wave traversing through the whole CH. In our simulations we assume a homogenous magnetic field which does not reflect the actual magnetic field structure of a CH in the observations. The complexity of the magnetic field structure inside a CH may also be a cause for the wave not being transmitted through the CH due to, \eg\ , dispersion of the wave on inhomogenities. We also have to be aware that our simulations are restricted to two dimensions, that is, the wave front is not capable of moving in the vertical direction as it would be the case in the observations.

\section{Conclusions}

We present the results of a newly developed 2.5D MHD code performing simulations of a fast mode MHD wave interacting with CHs of different density and various Alfv\'{e}n speed, respectively. In \citet{Piantschitsch2017} we demonstrated that the impact of the incoming wave causes different effects like reflection, transmission and the formation of stationary fronts for the case of an initial density amplitude of $\rho=1.5$ and a fixed initial CH density of $\rho_{CH}=0.1$. 

In this paper, we focus on comparing the properties of the different secondary waves and the stationary features with regard to various CH densities and different Alfv\'{e}n speed, respectively. We observe that the CH density is correlated to the amplitude values of the secondary waves and the peak values of the stationary features. The main simulation results look as follows:

\begin{itemize}
\item For the first traversing wave we found that the smaller the initial CH density, the smaller the wave's density amplitude and magnetic field component in $z$-direction, and the larger the amplitudes for phase speed and plasma flow velocity (see Figure \ref{morphology_IN_CH_no0} and Figure \ref{Kin_traver_wave}). The crucial point is, that the different CH densities correspond to different Alfv\'{e}n speeds inside the CH and hence to different phase speeds of the traversing waves. 
\item The analysis of the transmitted waves showed that the smaller the initial CH density, the smaller the amplitudes for density, magnetic field component in the $z$-direction and plasma flow velocity, and the larger the phase speed (see Figure \ref{zoom_transmitted_wave} and Figure \ref{Kin_transm_wave}).
\item We observe a very weak dependence of the first reflection on the CH density with regard to the initial parameters we choose for our simulations. The reflection seems to be mostly driven by the impact of the incoming wave on the CH boundary (see Figure \ref{morphology_1D_part2} and Figure \ref{Kin_first_reflection}).
\item The kinematic analysis of the second reflection has shown that we do not find a linear correlation between the initial CH density and the peak values for the different parameters of this feature like we have found for traversing and transmitted wave as well as for both stationary features (see Figure \ref{Kin_second_reflection}).
\item For the first stationary feature we have demonstrated that the smaller the initial CH density, the smaller the peak values of density and of magnetic field component in the $z$-direction. The stationary feature is moving slightly in the positive $x$-direction. (see Figure \ref{zoom_first_stat} and Figure \ref{Kin_first_stat})
\item On the contrary, in the case of the second stationary feature we observe that the smaller the initial CH density, the larger the peak values of density and magnetic field component. This second stationary feature is moving slightly in the negative $x$-direction (see Figure \ref{zoom_second_stat} and Figure \ref{Kin_second_stat}).
\item By analyzing the kinematics of the density depletion, we found that the smaller the initial CH density, the smaller the minimum density values of the depletion. Moreover, we find that the smaller the density values inside the CH, the larger the values of plasma flow velocity and phase speed (see Figure \ref{zoom_density_depletion} and Figure \ref{Kin_density_depletion}).
\end{itemize}

As already shown in \citet{Piantschitsch2017}, these findings strongly support the wave interpretation of large-scale disturbances in the corona. Firstly, effects like reflection and transmission can only be explained by a wave theory. We do not know of any other mechanism that would explain reflection or transmission of coronal waves. Secondly, the simulation results show that the interaction of an MHD wave and a CH is capable of forming stationary features, which were one of the main reasons for the development of a pseudo-wave theory. 

We compared our simulation results to observations in \citet{Kienreich_etal2012}, where the authors observed reflected features which consist of a bright lane followed by a dark lane in base-difference images. These observations correspond to the first reflection and the density depletion in our simulation.

\acknowledgments

The authors gratefully acknowledge the helpful comments from the anonymous referee that have very much improved the quality of this paper. This work was supported by the Austrian Science Fund (FWF): P23618 and P27765. B.V. acknowledges financial support by the Croatian Science Foundation under the project 6212 „Solar and Stellar Variability“. I.P. is grateful to Ewan C. Dickson for the proof read of this manuscript. The authors gratefully acknowledge support from NAWI Graz.

\bibliography{references}

\end{document}